\documentclass{article}
\usepackage{amssymb}
\usepackage{amsmath}
\usepackage{graphicx}

\newcommand{\va}{\mbox{\boldmath${a}$}}
\newcommand{\vg}{\mbox{\boldmath${g}$}}
\newcommand{\vk}{\mbox{\boldmath${k}$}}
\newcommand{\vn}{\mbox{\boldmath${n}$}}
\newcommand{\vp}{\mbox{\boldmath${p}$}}
\newcommand{\vv}{\mbox{\boldmath${v}$}}
\newcommand{\vx}{\mbox{\boldmath${x}$}}
\newcommand{\vA}{\mbox{\boldmath${A}$}}
\newcommand{\vF}{\mbox{\boldmath${F}$}}
\newcommand{\vJ}{\mbox{\boldmath${J}$}}
\newcommand{\vnabla}{\mbox{\boldmath${\nabla}$}}

\title{\bf \Large General Relativity Today\footnote{Talk given at the
Poincar\'e Seminar ``Gravitation et Exp\'erience''  (28 October 2006, Paris);
to appear in the proceedings to be published by Birkh\"auser.} \footnote{ 
Translated from the French by Eric Novak.}}
\author{Thibault Damour \\ \\
{\it Institut des Hautes Etudes Scientifiques} \\
{\it 35 route de Chartres, 91440 Bures-sur-Yvette, France}}
\date{ \ }

\begin{document}

\maketitle

\textbf{Abstract:} After recalling the conceptual foundations and 
the basic structure of general relativity, we review some of its main 
modern developments (apart from cosmology) : (i) the post-Newtonian
limit and weak-field tests in the solar system, (ii) strong gravitational
fields and black holes, (iii) strong-field and radiative  tests in binary pulsar
observations, (iv) gravitational waves, (v) general relativity and quantum theory.

\section{Introduction}\label{sec1}

The {\it theory of general relativity} was developed by Einstein
in work that extended from 1907 to 1915.  The starting point for
Einstein's thinking was the composition of a review article in
1907 on what we today call the {\it theory of special relativity}.
Recall that the latter theory sprang from a new kinematics
governing length and time measurements that was proposed by
Einstein in June of 1905 \cite{E05}, \cite{oeuvres}, following
important pioneering work by Lorentz and Poincar\'e. The theory of
special relativity essentially poses a new fundamental framework
(in place of the one posed by Galileo, Descartes, and Newton) for
the formulation of physical laws:  this framework being the
chrono-geometric
space-time structure of Poincar\'e and Minkowski. After 1905, it
therefore seemed a natural task to formulate, reformulate, or
modify the then known physical laws so that they fit within the
framework of special relativity. For Newton's law of gravitation,
this task was begun (before Einstein had even supplied his
conceptual crystallization in 1905) by Lorentz (1900) and
Poincar\'e (1905), and was pursued in the period from 1910 to 1915
by Max Abraham, Gunnar Nordstr\"om and Gustav Mie (with these
latter researchers developing {\it scalar} relativistic theories
of gravitation).

Meanwhile, in 1907, Einstein became aware that gravitational
interactions possessed particular characteristics that suggested
the necessity of {\it generalizing}  the framework and structure
of the 1905 theory of relativity. After many years of intense
intellectual effort, Einstein succeeded in constructing a {\it
generalized theory of relativity} (or {\it general relativity})
that proposed a profound modification of the chrono-geometric
structure of the space-time of special relativity. In 1915, in
place of a simple, neutral arena, given a priori, independently of
all material content,  space-time became a physical ``field''
(identified with the gravitational field).  In other words, it was
now a dynamical entity, both influencing and influenced by the
distribution of mass-energy that it contains.

This radically new conception of the structure of space-time
remained for a long while on the margins of the development of
physics. Twentieth century physics discovered a great number of
new physical laws and phenomena  while working with the space-time
of special relativity as its fundamental framework, as well as
imposing the respect of its symmetries (namely the
Lorentz-Poincar\'e group). On the other hand, the theory of
general relativity seemed for a long time to be a theory that was
both poorly confirmed by experiment and without connection to the
extraordinary progress springing from application of quantum
theory (along with special relativity) to high-energy physics.
This marginalization of general relativity no longer obtains.
Today, general relativity has become one of the essential players
in cutting-edge science.  Numerous high-precision experimental
tests have confirmed, in detail, the pertinence of this theory.
General relativity has become the favored tool for the description
of the macroscopic universe, covering everything from the big bang
to black holes, including the solar system, neutron stars,
pulsars, and gravitational waves. Moreover, the search for a
consistent description of fundamental physics in its entirety has
led to the exploration of theories that unify, within a general
quantum framework, the description of matter and all its
interactions (including gravity).  These theories, which are still
under construction and are provisionally known as string theories,
contain general relativity in a central way but suggest that the
fundamental structure of space-time-matter is even richer than is
suggested separately by quantum theory and general relativity.

\section{Special Relativity}\label{sec2}

We begin our exposition of the theory of general relativity by
recalling the chrono-geometric structure of space-time in the
theory of {\it special} relativity. The structure of
Poincar\'e-Minkowski
space-time is given by a generalization of the Euclidean geometric
structure of ordinary space. The latter structure is summarized by
the formula $L^2 = (\Delta x)^2 + (\Delta y)^2 + (\Delta z)^2$ (a
consequence of the Pythagorean theorem), expressing the square of
the distance $L$ between two points in space as  a sum of the
squares of the differences of the (orthonormal) coordinates
$x,y,z$ that label the points. The symmetry group of Euclidean
geometry is the group of coordinate transformations $(x,y,z) \to
(x',y',z')$ that leave the quadratic form $L^2 = (\Delta x)^2 +
(\Delta y)^2 + (\Delta z)^2$ invariant. (This group is generated
by translations, rotations, and ``reversals'' such as the
transformation given by reflection in a mirror, for example: $x' =
-x$, $y' = y$, $z'=z$.)

The Poincar\'e-Minkowski space-time is defined as the ensemble of
{\it events} (idealizations of what happens at a particular point
in space, at a particular moment in time), together with the
notion of a {\it (squared) interval} $S^2$ defined between any two
events. An event is fixed by four coordinates, $x,y,z$, and $t$,
where $(x,y,z)$ are the spatial coordinates of the point in space
where the event in question ``occurs,'' and where $t$ fixes the
instant when this event  ``occurs.''  Another event will be
described (within the same reference frame) by four different
coordinates, let us say $x + \Delta x$, $y + \Delta y$, $z +
\Delta z$, and $t + \Delta t$. The points in space where these two
events occur are separated by a distance $L$ given by the formula
above, $L^2 = (\Delta x)^2 + (\Delta y)^2 + (\Delta z)^2$. The
moments in time when these two events occur are separated by a
time interval $T$ given by $T = \Delta t$. The squared interval
$S^2$ between these two events is given as a function of these
quantities, by definition, through the following generalization of
the Pythagorean theorem:
\begin{equation}
\label{rg1} S^2 = L^2 - c^2 \, T^2 = (\Delta x)^2 + (\Delta y)^2 +
(\Delta z)^2 - c^2 (\Delta t)^2 \, ,
\end{equation}
where $c$ denotes the speed of light (or, more precisely, the
maximum speed of signal propagation).

Equation (\ref{rg1}) defines the {\it chrono-geometry} of
Poincar\'e-Minkowski space-time. The symmetry group of this
chrono-geometry is the group of coordinate transformations
$(x,y,z,t) \to (x',y',z',t')$ that leave the quadratic form
(\ref{rg1}) of the interval $S$ invariant. We will show that this
group is made up of linear transformations and that it is
generated by translations in space and time, spatial rotations,
``boosts'' (meaning special Lorentz transformations), and
reversals of space and time.

It is useful to replace the time coordinate $t$ by the
``light-time'' $x^0 \equiv ct$, and to collectively denote the
coordinates as $x^{\mu} \equiv (x^0 , x^i)$ where the Greek
indices $\mu , \nu , \ldots = 0,1,2,3$, and the Roman indices
$i,j,\ldots = 1,2,3$ (with $x^1 = x$, $x^2 = y$, and $x^3 = z$).
Equation (\ref{rg1}) is then written
\begin{equation}
\label{rg2}
S^2 = \eta_{\mu\nu} \, \Delta x^{\mu} \, \Delta x^{\nu} \, ,
\end{equation}
where we have used the Einstein summation
convention\footnote{Every repeated index is supposed to be summed
over all of its possible values.} and where $\eta_{\mu\nu}$ is a
diagonal matrix whose only non-zero elements are $\eta_{00} = -1$
and $\eta_{11} = \eta_{22} = \eta_{33} = +1$. The symmetry group
of Poincar\'e-Minkowski space-time is therefore the ensemble of
Lorentz-Poincar\'e transformations,
\begin{equation}
\label{rg3} x'^{\mu} = \Lambda_{\nu}^{\mu} \, x^{\nu} + a^{\mu} \,
,
\end{equation}
where $\eta_{\alpha \beta} \, \Lambda_{\mu}^{\alpha} \,
\Lambda_{\nu}^{\beta} = \eta_{\mu\nu}$.

The chrono-geometry of Poincar\'e-Minkowski space-time can be
visualized by representing, around each point $x$ in space-time,
the locus of points that are separated from the point $x$ by a
unit (squared) interval, in other words the ensemble of points
$x'$ such that $S_{xx'}^2 = \eta_{\mu\nu} (x'^{\mu} - x^{\mu})
(x'^{\nu} - x^{\nu}) = + 1$. This locus is a one-sheeted (unit)
hyperboloid.

\begin{figure}[h]
$$
\includegraphics[width=50mm]{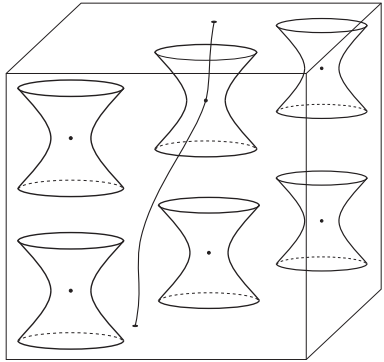}
$$
\caption{Geometry of the ``rigid'' space-time of the theory of
special relativity. This geometry is visualized by representing,
around each point $x$ in space-time, the locus of points separated
from the point $x$ by a unit (squared) interval. The space-time
shown here has only  three dimensions: one time dimension
(represented vertically), $x^0 = c t$, and two spatial dimensions
(represented horizontally), $x,y$. We have also shown the
`space-time line', or `world-line', (moving from the bottom to the top of the
``space-time block,'' or from the past towards the future)
representing the history of a particle's motion. }\label{fig1}
\end{figure}

If we were within an ordinary Euclidean space, the ensemble of
points $x'$ would trace out a (unit) sphere centered on $x$, and
the ``field'' of these spheres centered on each point $x$ would
allow one to completely characterize the Euclidean geometry of the
space. Similarly, in the case of Poincar\'e-Minkowski space-time,
the ``field'' of unit hyperboloids centered on each point $x$ is a
visual characterization of the geometry of this space-time. See
Figure~\ref{fig1}. This figure gives an idea of the symmetry group
of Poincar\'e-Minkowski space-time, and renders the rigid and
homogeneous nature of its geometry particularly clear.

The essential idea in Einstein's article of June 1905 was to
impose the group of transformations (\ref{rg3}) as a symmetry
group of the fundamental laws of physics (``the principle of
relativity''). This point of view proved to be extraordinarily
fruitful, since it led to the discovery of new laws and the
prediction of new phenomena.  Let us mention some of these for the
record: the relativistic dynamics of classical particles, the
dilation of lifetimes for relativistic particles, the relation $E
= mc^2$ between energy and inertial mass, Dirac's relativistic
theory of quantum ${\rm spin} \, \frac{1}{2}$ particles, the
prediction of antimatter, the classification of particles by rest
mass and spin, the relation between spin and statistics, and the
CPT theorem.

After these recollections on special relativity, let us discuss
the special feature of gravity
 which, in 1907, suggested to Einstein the need for a profound
generalization of the chrono-geometric structure of space-time.

\section{The Principle of Equivalence}\label{sec3}

Einstein's point of departure was a striking experimental fact:
all bodies in an external gravitational field fall with the same
acceleration. This fact was pointed out by Galileo in 1638.
Through a remarkable combination of logical reasoning, thought
experiments, and real experiments performed on inclined
planes,\footnote{The experiment with falling bodies said to be
performed from atop the Leaning Tower of Pisa is a myth, although
it aptly summarizes the essence of Galilean innovation.} Galileo
was in fact the first to conceive of what we today call the
``universality of free-fall'' or the ``weak principle of
equivalence.''  Let us cite the conclusion that Galileo drew from
a hypothetical argument where he varied the ratio between the
densities of the freely falling bodies under consideration and the
resistance of the medium through which they fall: ``Having
observed this I came to the conclusion that in a medium totally
devoid of resistance all bodies would fall with the same speed''
\cite{G70}. This universality of free-fall was verified with more
precision by Newton's experiments with pendulums, and was
incorporated by him into his theory of gravitation (1687) in the
form of the identification of the inertial mass $m_i$ (appearing
in the fundamental law of dynamics $\vF = m_i \, \va$) with the
gravitational mass $m_g$ (appearing in the gravitational force,
$F_g = G \, m_g \, m'_g / r^2$):
\begin{equation}
\label{rg4}
m_i = m_g \, .
\end{equation}

At the end of the nineteenth century, Baron Roland von E\"otv\"os
verified the equivalence (\ref{rg4}) between $m_i$ and $m_g$ with
a precision on the order of $10^{-9}$, and Einstein was aware of
this high-precision verification. (At present, the equivalence
between $m_i$ and $m_g$ has been verified at the level of
$10^{-12}$ \cite{tests}.) The point that struck Einstein was that,
given the precision with which $m_i = m_g$ was verified, and given
the equivalence between inertial mass and energy discovered by
Einstein in September of 1905 \cite{oeuvres} ($E = m_i \, c^2$),
one must conclude that all of the various forms of energy that
contribute to the inertial mass of a body (rest mass of the elementary
constituents, various binding energies, internal kinetic energy,
etc.) do contribute in a strictly identical way to the
gravitational mass of this body, meaning both to its capacity for
reacting to an external gravitational field and to its capacity to
create a gravitational field.

In 1907, Einstein realized that the equivalence between $m_i$ and
$m_g$ implicitly contained a deeper equivalence between inertia
and gravitation that had important consequences for the notion of
an inertial reference frame (which was a fundamental concept in
the theory of special relativity). In an ingenious thought
experiment, Einstein imagined the behavior of rigid bodies
and reference clocks within a freely falling elevator. Because of
the universality of free-fall, all of the objects in such a
``freely falling local reference frame'' would appear not to be
accelerating with respect to it. Thus, with respect to such a
reference frame, the exterior gravitational field is ``erased''
(or ``effaced'').
Einstein therefore postulated what he called the ``principle of
equivalence'' between gravitation and inertia. This principle has
two parts, that Einstein used in turns. The first part says that,
for any external gravitational field whatsoever, it is possible to
locally ``erase'' the gravitational field by using an appropriate
freely falling local reference frame and that, because of this,
the non-gravitational physical laws apply within this local
reference frame just as they would in an inertial reference frame
(free of gravity) in special relativity.  The second part of
Einstein's equivalence principle says that, by starting from an
inertial reference frame in special relativity (in the absence of
any ``true'' gravitational field), one can create an apparent
gravitational field in a local reference frame, if this reference
frame is accelerated (be it in a straight line or through a
rotation).

\section{Gravitation and Space-Time Chrono-Geometry}\label{sec4}

Einstein was able (through an extraordinary intellectual journey
that lasted eight years) to construct a new theory of gravitation,
based on a rich generalization of the 1905 theory of relativity,
starting just from the equivalence principle described above.  The
first step in this journey consisted in understanding that the
principle of equivalence would suggest a profound modification of
the chrono-geometric structure of Poincar\'e-Minkowski space-time
recalled in Equation (\ref{rg1}) above. 

To illustrate, let
$X^{\alpha}$, $\alpha = 0,1,2,3$, be the space-time coordinates in
a local, freely-falling reference frame (or {\it locally inertial
reference frame}). In such a reference frame, the laws of special
relativity apply. In particular, the infinitesimal space-time
interval $ds^2 = dL^2 - c^2 \, dT^2$ between two neighboring
events within such a reference frame $X^{\alpha}$, $X'^{\alpha} =
X^{\alpha} + dX^{\alpha}$ (close to the center of this reference
frame) takes the form
\begin{equation}
\label{rg5} ds^2 = dL^2 - c^2 \, dT^2 = \eta_{\alpha\beta} \,
dX^{\alpha} \, dX^{\beta} \, ,
\end{equation}
where we recall that the repeated indices $\alpha$ and $\beta$ are
summed over all of their values ($\alpha , \beta = 0,1,2,3$). We
also know that in special relativity the local energy and momentum
densities and fluxes are collected into the ten components of the
{\it energy-momentum tensor} $T^{\alpha\beta}$. (For example, the
energy density per unit volume is equal to $T^{00}$, in the
reference frame described by coordinates $X^{\alpha} = (X^0 ,
X^i)$, $i=1,2,3$.) The conservation of energy and momentum
translates into the equation $\partial_{\beta} \, T^{\alpha\beta}
= 0$, where $\partial_{\beta} = \partial / \partial \, X^{\beta}$.

The theory of special relativity tells us that we can change our
locally inertial reference frame (while remaining in the
neighborhood of a space-time point where one has ``erased''
gravity) through a Lorentz transformation, $X'^{\alpha} =
\Lambda_{\beta}^{\alpha} \, X^{\beta}$. Under such a
transformation, the infinitesimal interval $ds^2$, Equation
(\ref{rg5}), remains invariant and the ten components of the
(symmetric) tensor $T^{\alpha\beta}$ are transformed according to
$T'^{\alpha\beta} = \Lambda^{\alpha}_{\gamma} \,
\Lambda^{\beta}_{\delta} \, T^{\gamma \delta}$. On the other hand,
when we pass from a {\it locally} inertial reference frame (with
coordinates  $X^{\alpha}$) to an {\it extended} non-inertial
reference frame (with coordinates $x^{\mu}$; $\mu = 0,1,2,3$), the
transformation connecting the $X^{\alpha}$ to the $x^{\mu}$ is no
longer a {\it linear} transformation (like the Lorentz
transformation) but becomes a {\it non-linear} transformation
$X^{\alpha} = X^{\alpha} (x^{\mu})$ that can take any form
whatsoever. Because of this, the value of the infinitesimal
interval $ds^2$, when expressed in a general, extended reference
frame, will take a more complicated form than the very simple one
given by  Equation~(\ref{rg5}) that it had in a reference frame
that was locally in free-fall. In fact, by differentiating the
non-linear functions $X^{\alpha} = X^{\alpha} (x^{\mu})$ we obtain
the relation $dX^{\alpha} = \partial X^{\alpha} / \partial x^{\mu}
\, dx^{\mu}$. By substituting this relation into (\ref{rg5}) we
then obtain
\begin{equation}
\label{rg6}
ds^2 = g_{\mu\nu} (x^{\lambda}) \, dx^{\mu} \, dx^{\nu} \, ,
\end{equation}
where the indices $\mu , \nu$ are summed over $0,1,2,3$ and where
the ten functions $g_{\mu\nu} (x)$ (symmetric over the indices
$\mu$ and $\nu$) of the four variables $x^{\lambda}$ are defined,
point by point (meaning that for each point $x^{\lambda}$ we
consider a reference frame that is locally freely falling at $x$,
with local coordinates $X_x^{\alpha}$) by $g_{\mu\nu} (x) =
\eta_{\alpha\beta} \, \partial X_x^{\alpha} (x) / \partial x^{\mu}
\, \partial X_x^{\beta} (x) / \partial x^{\nu}$. Because of the
nonlinearity of the functions $X^{\alpha} (x)$, the functions
$g_{\mu\nu} (x)$ generally depend in a nontrivial way on the
coordinates  $x^{\lambda}$.

The local chrono-geometry of space-time thus appears to be given,
not by the simple Minkowskian metric (\ref{rg2}), with constant
coefficients $\eta_{\mu\nu}$, but by a quadratic metric of a much
more general type, Equation~(\ref{rg6}), with coefficients
$g_{\mu\nu} (x)$ that vary from point to point. Such general
metric spaces had been introduced and studied by Gauss and Riemann
in the nineteenth century (in the case where the quadratic form
(\ref{rg6}) is positive definite). They carry the name {\it
Riemannian spaces} or {\it curved spaces}. (In the case of
interest for Einstein's theory, where the quadratic form
(\ref{rg6}) is not positive definite, one speaks of a
pseudo-Riemannian metric.)

We do not have the space here to explain in detail the various
geometric structures in a Riemannian space that are derivable from
the data of the infinitesimal interval (\ref{rg6}). Let us note
simply that given Equation~(\ref{rg6}), which gives the distance
$ds$ between two infinitesimally separated points, we are able,
through integration along a curve, to define the length of an
arbitrary curve connecting two widely separated points $A$ and
$B$: $L_{AB} = \int_A^B ds$. One can then define the ``straightest
possible line'' between two given points $A$ and $B$ to be the
shortest line, in other words the curve that minimizes (or, more
generally, extremizes) the integrated distance $L_{AB}$. These
straightest possible lines are called {\it geodesic curves}. To
give a simple example, the geodesics of a spherical surface (like
the surface of the Earth) are the great circles (with radius equal
to the radius of the sphere). If one mathematically writes the
condition for a curve, as given by its parametric representation
$x^{\mu} = x^{\mu} (s)$, where $s$ is the length along the curve,
to extremize the total length $L_{AB}$ one finds that $x^{\mu}
(s)$ must satisfy the following second-order differential
equation:
\begin{equation}
\label{rg7}
\frac{d^2 \, x^{\lambda}}{ds^2} + \Gamma_{\mu\nu}^{\lambda} (x) \,
\frac{dx^{\mu}}{ds} \, \frac{dx^{\nu}}{ds} = 0 \, ,
\end{equation}
where the quantities $ \Gamma_{\mu\nu}^{\lambda}$, known as the
{\it Christoffel coefficients } or {\it connection coefficients},
are calculated, at each point $x$, from the {\it metric
components} $g_{\mu\nu} (x)$ by the equation
\begin{equation}
\label{rg8}
\Gamma_{\mu\nu}^{\lambda} \equiv \frac{1}{2} \, g^{\lambda\sigma}
(\partial_{\mu} \, g_{\nu\sigma} + \partial_{\nu} \, g_{\mu\sigma} -
\partial_{\sigma} \, g_{\mu\nu}) \, ,
\end{equation}
where $g^{\mu\nu}$ denotes the matrix inverse to $g_{\mu\nu}$
($g^{\mu\sigma} \, g_{\sigma\nu} = \delta_{\nu}^{\mu}$ where the
Kronecker symbol $\delta_{\nu}^{\mu}$ is equal to $1$ when $\mu =
\nu$ and $0$ otherwise) and where $\partial_{\mu} \equiv \partial
/ \partial x^{\mu}$ denotes the partial derivative with respect to
the coordinate $x^{\mu}$. To give a very simple example: in the
Poincar\'e-Minkowski space-time the components of the metric are
constant, $g_{\mu\nu} = \eta_{\mu\nu}$ (when we use an inertial
reference frame). Because of this, the connection coefficients
(\ref{rg8}) vanish in an inertial reference frame, and the
differential equation for geodesics reduces to $d^2 \, x^{\lambda}
/ ds^2 = 0$, whose solutions are ordinary straight lines:
$x^{\lambda} (s) = a^{\lambda} \, s + b^{\lambda}$. On the other
hand, in a general ``curved'' space-time (meaning one with
components $g_{\mu\nu}$ that depend in an arbitrary way on the
point $x$) the geodesics cannot be {\it globally} represented by
straight lines. One can nevertheless show that it always remains
possible, for any $g_{\mu\nu} (x)$ whatsoever, to change
coordinates $x^{\mu} \to X^{\alpha} (x)$ in such a way that the
connection coefficients $\Gamma_{\beta\gamma}^{\alpha}$, in the
new system of coordinates $X^{\alpha}$, vanish {\it locally}, at a
given point $X_0^{\alpha}$ (or even along an arbitrary curve).
Such {\it locally geodesic} coordinate systems realize Einstein's
equivalence principle mathematically: up to terms of second order,
the components $g_{\alpha\beta} (X)$ of a ``curved'' metric in
locally geodesic coordinates $X^{\alpha}$ ($ds^2 = g_{\alpha\beta}
(X) \, dX^{\alpha} \, dX^{\beta}$) can be identified with the
components of a ``flat''  Poincar\'e-Minkowski metric:
$g_{\alpha\beta} (X) = \eta_{\alpha\beta} + {\mathcal O}
((X-X_0)^2)$, where $X_0$ is the point around which we expand.

\section{Einstein's Equations: Elastic Space-Time}\label{sec5}

Having postulated that a consistent relativistic theory of the
gravitational field should include the consideration of a
far-reaching generalization of the Poincar\'e-Minkowski
space-time, Equation~(\ref{rg6}), Einstein concluded that the same
ten functions $\vg_{\mu\nu} (x)$ should describe both the {\bf
g}eometry of space-time as well as {\bf g}ravitation. He therefore
got down to the task of finding which equations must be satisfied
by the ``geometric-gravitational field'' $g_{\mu\nu} (x)$. He was
guided in this search by three principles.  The first was the {\it
principle of general relativity}, which asserts that in the
presence of a gravitational field one should be able to write
the fundamental laws of physics
(including those governing the gravitational field itself) 
in the same way in any coordinate system
whatsoever. The second was that the ``source'' of the
gravitational field should be the energy-momentum tensor
$T^{\mu\nu}$. The third was a principle of {\it correspondence}
with earlier physics: in the limit where one neglects
gravitational effects, $g_{\mu\nu} (x) = \eta_{\mu\nu}$ should be
a solution of the equations being sought, and there should also be
a so-called {\it Newtonian} limit where the new theory reduces to
Newton's theory of gravity.

Note that the principle of general relativity (contrary to
appearances and contrary to what Einstein believed for several
years) has a different physical status than the principle of
special relativity. The principle of special relativity was a
symmetry principle for the structure of space-time that asserted
that physics is {\it the same} in a particular class of reference
frames, and therefore that certain ``corresponding'' phenomena
occur in exactly the same way in different reference frames
(``active'' transformations). On the other hand, the principle of
general relativity is a {\it principle of indifference}: the
phenomena do not (in general) take place in the same way in
different coordinate systems.  However, none of these (extended)
coordinate systems enjoys any privileged status with respect to
the others.

The principle asserting that the energy-momentum tensor
$T^{\mu\nu}$ should be the source of the gravitational field is
founded on two ideas: the relations $E = m_i \, c^2$ and the weak
principle of equivalence $m_i = m_g$ show that, in the Newtonian
limit, the source of gravitation, the gravitational mass $m_g$, is
equal to the total energy of the body considered, or in other
words the integral over space of the energy density $T^{00}$, up
to the factor $c^{-2}$. Therefore at least one of the components
of the tensor $T^{\mu\nu}$ must play the role of source for the
gravitational field. However, since the gravitational field is
encoded, according to Einstein, by the ten components of the
metric $g_{\mu\nu}$, it is natural to suppose that the source for
$g_{\mu\nu}$ must also have ten components, which is precisely the
case for the (symmetric) tensor $T^{\mu\nu}$.

In November of 1915, after many years of conceptually arduous
work, Einstein wrote the final form of the theory of general
relativity \cite{livres}.
{\it Einstein's equations} are non-linear, second-order partial
differential equations for the geometric-gravitational field
$g_{\mu\nu}$, containing the energy-momentum tensor $T_{\mu\nu}
\equiv g_{\mu\kappa} \, g_{\nu\lambda} \, T^{\kappa\lambda}$ on
the right-hand side. They are written as follows:
\begin{equation}
\label{rg9}
R_{\mu\nu} - \frac{1}{2} \, R \, g_{\mu\nu} = \frac{8\pi \, G}{c^4} \,
T_{\mu\nu}
\end{equation}
where $G$ is the (Newtonian) gravitational constant, $c$ is the
speed of light, and $R \equiv g^{\mu\nu} \, R_{\mu\nu}$ and the
{\it Ricci tensor} $R_{\mu\nu}$ are calculated as a function of
the connection coefficients $\Gamma_{\mu\nu}^{\lambda}$
(\ref{rg8}) in the following way:
\begin{equation}
\label{rg10}
R_{\mu\nu} \equiv \partial_{\alpha} \, \Gamma_{\mu\nu}^{\alpha} - \partial_{\nu}
\, \Gamma_{\mu\alpha}^{\alpha} + \Gamma_{\beta\alpha}^{\alpha} \,
\Gamma_{\mu\nu}^{\beta} - \Gamma_{\beta\nu}^{\alpha} \,
\Gamma_{\mu\alpha}^{\beta} \, .
\end{equation}

One can show that, in a four-dimensional space-time, the three
principles we have described previously uniquely determine
Einstein's equations (\ref{rg9}). It is nevertheless remarkable
that these equations may also be developed from points of view
that are completely different from the one taken by Einstein. For
example, in the 1960s various authors (in particular Feynman,
Weinberg and Deser; see references in \cite{tests}) showed that
Einstein's equations could be obtained from a purely {\it
dynamical} approach, founded on the consistency of interactions of
a long-range spin $2$ field, without making any appeal, as
Einstein had, to the {\it geometric} notions coming from
mathematical work on Riemannian spaces. Let us also note that if
we relax one of the principles described previously (as Einstein
did in 1917) we can find a generalization of Equation (\ref{rg9})
in which one adds the term $+ \, \Lambda \, g_{\mu\nu}$ to the
left-hand side, where $\Lambda$ is the so-called {\it cosmological
constant}. Such a modification was proposed by Einstein in 1917 in
order to be able to write down a globally homogeneous and {\it
stationary} cosmological solution. Einstein rejected this
additional term after work by Friedmann (1922) showed the
existence of {\it expanding} cosmological solutions of general
relativity and after the observational discovery (by Hubble in
1929) of the expanding motion of galaxies within the universe.
However, recent cosmological data have once again made this
possibility fashionable, although in the fundamental physics of
today one tends to believe that a term of the type $\Lambda \,
g_{\mu\nu}$ should be considered as a particular physical
contribution to the right-hand side of Einstein's equations (more
precisely, as the stress-energy tensor of the {\it vacuum},
$T_{\mu\nu}^V = - \frac{c^4}{8\pi G} \, \Lambda \, g_{\mu\nu}$),
rather than as a universal geometric modification of the left-hand
side.

\begin{figure}[h]
$$
\includegraphics[width=50mm]{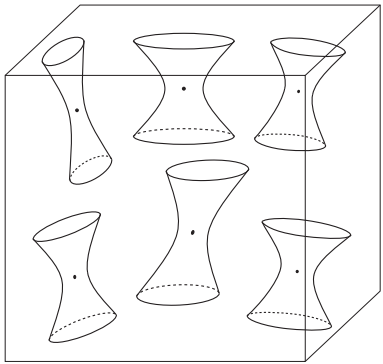}
$$
\caption{``Elastic'' space-time geometry in the theory of general
relativity. This geometry is visualized by representing, around
each space-time point $x$, the locus of points separated from $x$
by a given small positive (squared) interval.}\label{fig2}
\end{figure}

Let us now comment on the physical meaning of Einstein's equations
(\ref{rg9}). The essential new idea is that the chrono-geometric
structure of space-time, Equation~(\ref{rg6}), in other words the
structure that underlies all of the measurements that one could
locally make of duration, $dT$, and of distance, $dL$, (we recall
that, locally, $ds^2 = dL^2 - c^2 \, dT^2$) is no longer a rigid
structure that is given a priori, once and for all (as was the
case for the structure of Poincar\'e-Minkowski space-time), but
instead has become a {\it field}, a dynamical or {\it elastic}
structure, which is created and/or deformed by the presence of an
energy-momentum distribution. See Figure~\ref{fig2}, which
visualizes the ``elastic'' geometry of space-time in the theory of
general relativity by representing, around each point $x$, the
locus of points (assumed to be infinitesimally close to $x$)
separated from $x$ by a constant (squared) interval: $ds^2 =
\varepsilon^2$. As in the case of Poincar\'e-Minkowski space-time
(Figure~\ref{fig1}), one arrives at a ``field'' of hyperboloids.
 However, this field of hyperboloids no longer has a ``rigid'' and homogeneous structure.

The {\it space-time field} $g_{\mu\nu} (x)$ describes the
variation from point to point of the chrono-geometry as well as
all gravitational effects. The simplest example of space-time
chrono-geometric {\it elasticity} is the effect that the proximity
of a mass has on the ``local rate of flow for time.'' In concrete
terms, if you separate two twins at birth, with one staying on the
surface of the Earth and the other going to live on the peak of a
very tall mountain (in other words farther from the Earth's
center), and then reunite them after 100 years, the ``highlander''
will be older (will have lived longer) than the twin who stayed on
the valley floor. Everything takes place as if time flows more
slowly the closer one is to a given distribution of mass-energy.
In mathematical terms this effect is due to the fact that the
coefficient $g_{00} (x)$ of $(dx^0)^2$ in Equation~(\ref{rg6}) is
deformed with respect to its value in special relativity,
$g_{00}^{\rm Minkowski} = \eta_{00} = -1$, to become $g_{00}^{\rm
Einstein} (x) \simeq -1 + 2GM / c^2 r$, where $M$ is the Earth's
mass (in our example) and $r$ the distance to the center of the
Earth. In the example considered above of terrestrial twins the
effect is extremely small (a difference in the amount of time
lived of about one second over 100 years), but the effect is real
and has been verified many times using atomic clocks (see the
references in \cite{tests}). Let us mention that today this
``Einstein effect'' has important practical repercussions, for
example in aerial or maritime navigation, for the piloting of
automobiles, or even farm machinery, etc. In fact, the GPS (Global
Positioning System), which uses the data transmitted by a
constellation of atomic clocks on board satellites, incorporates
the Einsteinian deformation of space-time chrono-geometry into its
software.  The effect is only on the order of one part in a
billion, but if it were not taken into account, it would introduce
an unacceptably large error into the GPS, which would continually
grow over time. Indeed, GPS performance relies on the high
stability of the orbiting atomic clocks, a stability better than
$10^{-13}$, or in other words 10,000 times greater than the
apparent change in frequency($\sim 10^{-9}$) due to the
Einsteinian deformation of the chrono-geometry.

\section{The Weak-Field Limit and the Newtonian Limit}\label{sec6}

To understand the physical consequences of Einstein's equations
(\ref{rg9}), it is useful to begin by considering the limiting
case of {\it weak} geometric-gravitational fields, namely the case
where $g_{\mu\nu} (x) = \eta_{\mu\nu} + h_{\mu\nu} (x)$, with
perturbations $h_{\mu\nu} (x)$ that are very small with respect to
unity: $\vert h_{\mu\nu} (x) \vert \ll 1$. In this case, a simple
calculation (that we encourage the reader to perform) starting
from Definitions (\ref{rg8}) and (\ref{rg10}) above, leads to the
following explicit form of Einstein's equations (where we ignore
terms of order $h^2$ and $hT$):
\begin{equation}
\label{rg11}
\Box \, h_{\mu\nu} - \partial_{\mu} \, \partial^{\alpha} \, h_{\alpha \nu} -
\partial_{\nu} \, \partial^{\alpha} \, h_{\alpha\mu} + \partial_{\mu\nu} \,
h_{\alpha}^{\alpha} = - \frac{16 \, \pi \, G}{c^4} \, \tilde T_{\mu\nu} \, ,
\end{equation}
where $\Box = \eta^{\mu\nu} \, \partial_{\mu\nu} = \Delta -
\partial_0^2 =
\partial^2 / \partial x^2 +  \partial^2 / \partial y^2 +  \partial^2 / \partial
z^2 - c^{-2} \, \partial^2 / \partial t^2$ denotes the ``flat''
d'Alembertian (or wave operator; $x^{\mu} = (ct , x, y , z)$), and
where indices in the upper position have been raised by the
inverse $\eta^{\mu\nu}$ of the flat metric $\eta_{\mu\nu}$
(numerically $\eta^{\mu\nu} = \eta_{\mu\nu}$, meaning that
$-\eta^{00} = \eta^{11} = \eta^{22} = \eta^{33} = +1$). For
example $\partial^{\alpha} \, h_{\alpha\nu}$ denotes
$\eta^{\alpha\beta} \, \partial_{\alpha} \, h_{\beta\nu}$ and
$h_{\alpha}^{\alpha} \equiv \eta^{\alpha\beta} \, h_{\alpha\beta}
= -h_{00} + h_{11} + h_{22} + h_{33}$. The ``source'' $\tilde
T_{\mu\nu}$ appearing on the right-hand side of (\ref{rg11})
denotes the combination $\tilde T_{\mu\nu} \equiv T_{\mu\nu} -
\frac{1}{2} \, T_{\alpha}^{\alpha} \, \eta_{\mu\nu}$ (when
space-time is four-dimensional).

The ``linearized'' approximation (\ref{rg11}) of Einstein's
equations is analogous to Maxwell's equations
\begin{equation}
\label{rg12}
\Box \, A_{\mu} - \partial_{\mu} \, \partial^{\alpha} \, A_{\alpha} = -4\pi \,
J_{\mu} \, ,
\end{equation}
connecting the electromagnetic four-potential $A_{\mu} \equiv
\eta_{\mu\nu} \, A^{\nu}$ (where $A^0 = V$, $A^i = \vA$, $i =
1,2,3$) to the four-current density $J_{\mu} \equiv \eta_{\mu\nu}
\, J^{\nu}$ (where $J^0 = \rho$ is the charge density and $J^i =
\vJ$ is the current density). Another analogy is that the
structure of the left-hand side of Maxwell's equations implies
that the ``source'' $J_{\mu}$ appearing on the right-hand side
must satisfy $\partial^{\mu} \, J_{\mu} = 0$ ($\partial^{\mu}
\equiv \eta^{\mu\nu} \,
\partial_{\nu}$), which expresses the conservation of electric charge.
Likewise, the structure of the left-hand side of the linearized
form of Einstein's equations (\ref{rg11}) implies that the
``source'' $T_{\mu\nu} = \tilde T_{\mu\nu} - \frac{1}{2} \, \tilde
T_{\alpha}^{\alpha} \, \eta_{\mu\nu}$ must satisfy $\partial^{\mu}
\, T_{\mu\nu} = 0$, which expresses the conservation of energy and
momentum of matter. (The structure of the left-hand side of the
exact form of Einstein's equations (\ref{rg9}) implies that the
source $T_{\mu\nu}$ must satisfy the more complicated equation
$\partial_{\mu} \, T^{\mu\nu} + \Gamma_{\sigma\mu}^{\mu} \,
T^{\sigma\nu} + \Gamma_{\sigma\mu}^{\nu} \, T^{\mu\sigma} = 0$,
where the terms in $\Gamma T$ can be interpreted as describing an
exchange of energy and momentum between matter and the
gravitational field.) The major difference is that, in the case of
electromagnetism, the field $A_{\mu}$ and its source $J_{\mu}$
have a single space-time index, while in the gravitational case
the field $h_{\mu\nu}$ and its source $\tilde T_{\mu\nu}$ have two
space-time indices. We shall return later to this
analogy/difference between $A_{\mu}$ and $h_{\mu\nu}$ which
suggests the existence of a certain relation between gravitation
and electromagnetism.

We recover the Newtonian theory of gravitation as the limiting
case of Einstein's theory by assuming not only that the
gravitational field is a weak deformation of the flat Minkowski
space-time ($h_{\mu\nu} \ll 1$), but also that the field
$h_{\mu\nu}$ is slowly varying ($\partial_0 \, h_{\mu\nu} \ll
\partial_i \, h_{\mu\nu}$) and that its source $T_{\mu\nu}$ is
non-relativistic ($T_{ij} \ll T_{0i} \ll T_{00}$). Under these
conditions Equation (\ref{rg11}) leads to a Poisson-type equation
for the purely temporal component, $h_{00}$, of the space-time
field,
\begin{equation}
\label{rg13}
\Delta \, h_{00} = - \frac{16 \, \pi \, G}{c^4} \, \tilde T_{00} = - \frac{8 \,
\pi \, G}{c^4} \, (T_{00} + T_{ii}) \simeq - \frac{8 \, \pi \, G}{c^4} \, T_{00}
\, ,
\end{equation}
where $\Delta = \partial_x^2 + \partial_y^2 + \partial_z^2$ is the
Laplacian. Recall that, according to Laplace and Poisson, Newton's
theory of gravity is summarized by saying that the gravitational
field is described by a single potential $U(x)$, produced by the
mass density $\rho (x)$ according to the Poisson equation $\Delta
U = - 4 \, \pi \, G \rho$, that determines the acceleration of a
test particle placed in the exterior field $U(x)$ according to the
equation $d^2 \, x^i / dt^2 =
\partial_i \, U(x) \equiv
\partial U / \partial x^i$. Because of the relation $m_i = m_g = E/c^2$
one can identify $\rho = T^{00} / c^2$. We therefore find that
(\ref{rg13}) reproduces the Poisson equation if $h_{00} = + \, 2
\, U / c^2$. It therefore remains to verify that Einstein's theory
indeed predicts that a non-relativistic test particle is
accelerated by a space-time field according to $d^2 \, x^i / dt^2
\simeq \frac{1}{2} \, c^2 \, \partial_i \, h_{00}$. Einstein
understood that this was a consequence of the equivalence
principle. In fact,  as we discussed in Section~\ref{sec4}
above, the principle of equivalence states that the gravitational
field is (locally) erased in a locally inertial reference frame
$X^{\alpha}$ (such that $g_{\alpha\beta} (X) = \eta_{\alpha\beta}
+ {\mathcal O} ((X-X_0)^2)$). In such a reference frame, the laws
of special relativity apply at the point $X_0$. In particular an
isolated (and electrically neutral) body must satisfy a principle
of inertia in this frame: its center of mass moves in a straight
line at constant speed. In other words it satisfies the equation
of motion $d^2 \, X^{\alpha} / ds^2 = 0$. By passing back to an
arbitrary (extended) coordinate system $x^{\mu}$, one verifies
that this equation for inertial motion transforms into the
geodesic equation (\ref{rg7}). Therefore (\ref{rg7}) describes
falling bodies, such as they are observed in arbitrary extended
reference frames (for example a reference frame at rest with
respect to the Earth or at rest with respect to the center of mass
of the solar system). From this one concludes that the
relativistic analog of the Newtonian field of gravitational
acceleration, $\vg (x) = \vnabla \, U(x)$, is $g^{\lambda} (x)
\equiv - c^2 \, \Gamma_{\mu\nu}^{\lambda} \, dx^{\mu} /ds \,
dx^{\nu} / ds$. By considering a particle whose motion is slow
with respect to the speed of light ($dx^i / ds \ll dx^0 / ds
\simeq 1$) one can easily verify that $g^i (x) \simeq -c^2 \,
\Gamma_{00}^i$. Finally, by using the definition (\ref{rg8}) of
$\Gamma_{\mu\nu}^{\alpha}$, and the hypothesis of weak fields, one
indeed verifies that $g^i (x) \simeq \frac{1}{2} \, c^2 \,
\partial_i \, h_{00}$, in perfect agreement with the identification
$h_{00} = 2 \, U/c^2$ anticipated above. We encourage the reader
to personally verify this result, which contains the very essence
of Einstein's theory: gravitational motion is no longer described
as being due to a force, but is identified with motion that is
``as inertial as possible'' within a space-time whose
chrono-geometry is deformed in the presence of a mass-energy
distribution.

Finding the Newtonian theory as a limiting case of Einstein's
theory is obviously a necessity for seriously considering this new
theory. But of course, from the very beginning Einstein explored
the observational consequences of general relativity that go
beyond the Newtonian description of gravitation. We have already
mentioned one of these above: the fact that $g_{00} = \eta_{00} +
h_{00} \simeq -1 + 2 U (x) / c^2$ implies a distortion in the
relative measurement of time in the neighborhood of massive
bodies. In 1907 (as soon as he had developed the principle of
equivalence, and long before he had obtained the field equations
of general relativity) Einstein had predicted the existence of
such a distortion for measurements of time and frequency in the
presence of an external gravitational field. He realized that this
should have observable consequences for the frequency, as observed
on Earth, of the spectral rays emitted from the surface of the
Sun. Specifically, a spectral ray of (proper local) frequency
$\nu_0$ emitted from a point $x_0$ where the (stationary)
gravitational potential takes the value $U (\vx_0)$ and observed
(via electromagnetic signals) at a point $x$ where the potential
is $U(\vx)$ should appear to have a frequency $\nu$ such that
\begin{equation}
\label{rg14}
\frac{\nu}{\nu_0} = \sqrt{\frac{g_{00} (x_0)}{g_{00} (x)}} \simeq 1 +
\frac{1}{c^2} \, [U(\vx) - U (\vx_0)] \, .
\end{equation}
In the case where the point of emission $x_0$ is in a
gravitational potential well deeper than the point of observation
$x$ (meaning that $U (\vx_0)
> U(\vx)$) one has $\nu < \nu_0$, in other words a {\it reddening}
effect on frequencies. This effect, which was predicted by
Einstein in 1907, was unambiguously verified only in the 1960s, in
experiments by Pound and collaborators over a height of about
twenty meters. The most precise verification (at the level of
$\sim 10^{-4}$) is due to Vessot and collaborators, who compared a
hydrogen maser, launched aboard a rocket that reached about 10,000
km in altitude, to a clock of similar construction on the ground.
Other experiments compared the times shown on clocks placed aboard
airplanes to clocks remaining on the ground. (For references to
these experiments see \cite{tests}.) As we have already mentioned,
the ``Einstein effect'' (\ref{rg14}) must be incorporated in a
crucial way into the software of satellite positioning systems
such as the GPS.

In 1907, Einstein also pointed out that the equivalence principle
would suggest that light rays should be deflected by a
gravitational field. Indeed, a generalization of the reasoning
given above for the motion of particles in an external
gravitational field, based on the principle of equivalence,
shows that light must itself follow a trajectory that is ``as
inertial as possible,'' meaning a geodesic of the curved
space-time. Light rays must therefore satisfy the geodesic
equation (\ref{rg7}). (The only difference from the geodesics
followed by material particles is that the parameter $s$ in
Equation (\ref{rg7}) can no longer be taken equal to the
``length'' along the geodesic, since a ``light'' geodesic must
also satisfy the constraint $g_{\mu\nu} (x) \, dx^{\mu} \,
dx^{\nu} = 0$, ensuring that its speed is equal to $c$, when it is
measured in a locally inertial reference frame.) Starting from
Equation (\ref{rg7}) one can therefore calculate to what extent
light is deflected when it passes through the neighborhood of a
large mass (such as the Sun). One nevertheless soon realizes that
in order to perform this calculation one must know more than the
component $h_{00}$ of the gravitational field. The other
components of $h_{\mu\nu}$, and in particular the spatial
components $h_{ij}$, come into play in a crucial way in this
calculation. This is why it was only in November of 1915, after
having obtained the (essentially) final form of his theory, that
Einstein could predict the total value of the deflection of light
by the Sun. Starting from the linearized form of Einstein's
equations (\ref{rg11}) and continuing by making the
``non-relativistic'' simplifications indicated above ($T_{ij} \ll
T_{0i} \ll T_{00}$, $\partial_0 \, h \ll
\partial_i \, h$) it is easy to see that the spatial component
$h_{ij}$, like $h_{00}$, can be written (after a helpful choice of
coordinates) in terms of the Newtonian potential $U$ as $h_{ij}
(x) \simeq + \, 2 \, U(x) \, \delta_{ij} / c^2$, where
$\delta_{ij}$ takes the value $1$ if $i=j$ and $0$ otherwise ($i,j
= 1,2,3$). By inserting this result, as well as the preceding
result $h_{00} = + \, 2 \, U / c^2$, into the geodesic equation
(\ref{rg7}) for the motion of light, one finds (as Einstein did in
1915) that general relativity predicts that the Sun should deflect
a ray of light by an angle $\theta = 4GM / (c^2 b)$ where $b$ is
the impact parameter of the ray (meaning its minimum distance from
the Sun). As is well known, the confirmation of this effect in
1919 (with rather weak precision) made the theory of general
relativity and its creator famous.

\section{The Post-Newtonian Approximation and Experimental Confirmations
in the Regime of Weak and Quasi-Stationary Gravitational
Fields}\label{sec7}

We have already pointed out some of the experimental confirmations
of the theory of general relativity. At present, the extreme
precision of certain measurements of time or frequency in
the solar system necessitates a very careful account of the
modifications brought by general relativity to the Newtonian
description of space-time. As a consequence, general relativity is
used in a great number of situations, from astronomical or
geophysical research (such as very long range radio
interferometry, radar tracking of the planets, and laser tracking
of the Moon or artificial satellites) to metrological, geodesic or
other applications (such as the definition of international atomic
time, precision cartography, and the G.P.S.). To do this, the
so-called {\it post-Newtonian} approximation has been developed.
This method involves working in the Newtonian limit sketched above
while keeping the terms of higher order in the small parameter
$$
\varepsilon \sim \frac{v^2}{c^2} \sim \vert h_{\mu\nu} \vert \sim \vert
\partial_0 \, h / \partial_i \, h \vert^2 \sim \vert T^{0i} / T^{00} \vert^2
\sim \vert T^{ij} / T^{00} \vert \, ,
$$
where $v$ denotes a characteristic speed for the elements in the
system considered.

For all present applications of general relativity to the solar
system it suffices to include the {\it first post-Newtonian
approximation}, in other words to keep the relative corrections of
order $\varepsilon$ to the Newtonian predictions. Since the theory
of general relativity was poorly verified for a long time, one
found it useful (as in the pioneering work of A.~Eddington,
generalized in the 1960s by K.~Nordtvedt and C.M.~Will) to study
not only the precise predictions of the equations (\ref{rg9})
defining Einstein's theory, but to also consider possible
deviations from these predictions. These possible deviations were
parameterized by means of several non-dimensional
``post-Newtonian'' parameters.  Among these parameters, two play a
key role: $\gamma$ and $\beta$. The parameter $\gamma$ describes a
possible deviation from general relativity that comes into play
starting at the linearized level, in other words one that modifies
the linearized approximation given above. More precisely, it is
defined by writing that the difference $h_{ij} \equiv g_{ij} -
\delta_{ij}$ between the spatial metric and the Euclidean metric
can take the value $h_{ij} = 2 \gamma \, U \, \delta_{ij} / c^2$
(in a suitable coordinate system), rather than the value
$h_{ij}^{\rm GR} = 2 \, U \, \delta_{ij} / c^2$ that it takes in
general relativity, thus differing by a factor $\gamma$.
Therefore, by definition $\gamma$ takes the value $1$ in general
relativity, and $\gamma -1$ measures the possible deviation with
respect to this theory. As for the parameter $\beta$ (or rather
$\beta - 1$), it measures a possible deviation (with respect to
general relativity) in the value of $h_{00} \equiv g_{00} -
\eta_{00}$. The value of $h_{00}$ in general relativity is
$h_{00}^{\rm GR} = 2 \, U / c^2 - 2 \, U^2 / c^4$, where the first
term (discussed above) reproduces the Newtonian approximation (and
cannot therefore  be modified, as the idea is to parameterize
gravitational physics beyond Newtonian predictions) and where the
second term is obtained by solving Einstein's equations
(\ref{rg9}) at the second order of approximation. One then writes
an $h_{00}$ of a more general parameterized type, $h_{00} = 2 \, U
/ c^2 - 2 \, \beta \, U^2 / c^4$, where, by definition, $\beta$
takes the value $1$ in general relativity. Let us finally point
out that the parameters $\gamma - 1$ and $\beta - 1$ completely
parameterize the post-Newtonian regime of the simplest theoretical
alternatives to general relativity, namely the tensor-scalar
theories of gravitation. In these theories, the gravitational
interaction is carried by two fields at the same time: a massless
tensor (spin 2) field coupled to $T^{\mu\nu}$, and a massless
scalar (spin 0) field $\varphi$ coupled to the trace
$T_{\alpha}^{\alpha}$. In this case the parameter $-(\gamma -1)$
plays the key role of measuring the ratio between the scalar
coupling and the tensor
coupling.

All of the experiments performed to date within the solar system
are compatible with the predictions of general relativity. When
they are interpreted in terms of the post-Newtonian (and
``post-Einsteinian'') parameters $\gamma - 1$ and $\beta - 1$,
they lead to strong constraints on possible deviations from
Einstein's theory. We make note of the following among tests
performed in the solar system: the deflection of electromagnetic
waves in the neighborhood of the Sun, the gravitational delay 
(`Shapiro effect') of radar signals bounced from the Viking lander on
Mars, the global analysis of solar system dynamics (including the
advance of planetary perihelia), the sub-centimeter measurement of
the Earth-Moon distance obtained from laser signals bounced off of
reflectors on the Moon's surface, etc. At present (October of
2006) the most precise test (that has been published) of general
relativity was obtained in 2003 by measuring the ratio $1+y \equiv
f/f_0$ between the frequency $f_0$ of radio waves sent from Earth
to the Cassini space probe and the frequency $f$ of coherent radio
waves sent back (with the same local frequency) from Cassini to
Earth and compared (on Earth) to the emitted frequency $f_0$. The
main contribution to the small quantity $y$ is an effect equal, in
general relativity, to $y_{GR} = 8(GM/c^3 \, b) \, db/dt$ (where
$b$ is, as before, the impact parameter) due to the propagation of
radio waves in the geometry of a space-time deformed by the Sun:
$ds^2 \simeq - (1-2 \, U / c^2) \, c^2 \, dt^2 + (1+2 \, U / c^2)
(dx^2 + dy^2 + dz^2)$, where $U = GM/r$. The maximum value of the
frequency change predicted by general relativity was only $\vert
y_{\rm GR} \vert \lesssim 2 \times 10^{-10}$ for the best
observations, but thanks to an excellent frequency stability $\sim
10^{-14}$ (after correction for the perturbations caused by the
solar corona) and to a relatively large number of individual
measurements spread over 18 days, this experiment was able to
verify Einstein's theory at the remarkable level of $\sim 10^{-5}$
\cite{Cassini}. More precisely, when this experiment is
interpreted in terms of the post-Newtonian parameters $\gamma - 1$
and $\beta - 1$, it gives the following limit for the parameter
$\gamma - 1$ \cite{Cassini}
\begin{equation}
\label{rg15}
\gamma - 1 = (2.1 \pm 2.3) \times 10^{-5} \, .
\end{equation}
As for the best present-day limit on the parameter $\beta-1$, it
is smaller than $10^{-3}$ and comes from the non-observation, in
the data from lasers bounced off of the Moon, of any eventual
polarization of the Moon's orbit in the direction of the Sun 
(`Nordtvedt effect'; see \cite{tests} for
references)
\begin{equation}
\label{rg16}
4 (\beta - 1) - (\gamma - 1) = -0.0007 \pm 0.0010 \, .
\end{equation}

Although the theory of general relativity is one of the best
verified theories in physics, scientists continue to design and
plan new or increasingly precise tests of the theory. This is the
case in particular for the space mission Gravity Probe B (launched
by NASA in April of 2004) whose principal aim is to directly
observe a prediction of general relativity that states
(intuitively speaking) that space is not only ``elastic,'' but
also ``fluid.''  In the nineteenth century Foucalt invented both
the gyroscope and his famous pendulum in order to render Newton's
absolute (and rigid) space directly observable. His experiments in
fact showed that, for example, a gyroscope on the surface of the
Earth continued, despite the Earth's rotation, to align itself in
a direction that is ``fixed'' with respect to the distant stars.
However, in 1918, when Lense and Thirring analyzed some of the
consequences of the (linearized) Einstein equations (\ref{rg11}),
they found that general relativity predicts, among other things,
the following phenomenon: the rotation of the Earth (or any other
ball of matter) creates a particular deformation of the
chrono-geometry of space-time.  This deformation is described by
the ``gravito-magnetic'' components $h_{0i}$ of the metric, and
induces an effect analogous to the ``rotation drag'' effect caused
by a ball of matter turning in a fluid: the rotation of the Earth
(minimally) drags all of the space around it, causing it to
continually ``turn,'' as a fluid would.\footnote{Recent historical
work (by Herbert Pfister) has in fact shown that this effect had
already been derived by Einstein within the framework of the
provisory relativistic theory of gravity that he started to
develop in 1912 in collaboration with Marcel Grossmann.} This
``rotation of space'' translates, in an observable way, into a
violation of the effects predicted by Newton and confirmed by
Foucault's experiments: in particular, a gyroscope no longer aligns
itself in a direction that is ``fixed in absolute space,'' rather
its axis of rotation is ``dragged'' by the rotating motion of the
local space where it is located.  This effect is much too small to
be visible in Foucalt's experiments. Its observation by Gravity
Probe B (see \cite{gpb} and the contribution of John Mester to
this Poincar\'e seminar) is important for making Einstein's
revolutionary notion of a fluid space-time tangible to the general
public.

Up till now we have only discussed the regime of weak and slowly
varying gravitational fields.  The theory of general relativity
predicts the appearance of new phenomena when the gravitational
field becomes strong and/or rapidly varying. (We shall not here
discuss the cosmological aspects of relativistic gravitation.)

\section{Strong Gravitational Fields and Black Holes}\label{sec8}

The regime of strong gravitational fields is encountered in the
physics of {\it gravitationally condensed bodies}. This term
designates the final states of stellar evolution, and in
particular neutron stars and black holes.  Recall that most of the
life of a star is spent slowly burning its nuclear fuel. This
process causes the star to be structured as a series of layers of
differentiated nuclear structure, surrounding a progressively
denser core (an ``onion-like'' structure). When the initial mass
of the star is sufficiently large, this process ends into
 a catastrophic phenomenon: the core, already much
denser than ordinary matter, collapses in on itself under the
influence of its gravitational self-attraction. (This implosion of
the central part of the star is, in many cases, accompanied by an
explosion of the outer layers of the star---a supernova.)
Depending on the quantity of mass that collapses with the core of
a star, this collapse can give rise to either a neutron star or a
black hole.

A {\it neutron star} condenses a mass on the order of the mass of
the Sun inside a radius on the order of 10 km. The density in the
interior of a neutron star (named thus because neutrons dominate
its nuclear composition) is more than 100 million tons per cubic
centimeter ($10^{14}$ g/cm$^3$)! It is about the same as the
density in the interior of atomic nuclei. What is important for
our discussion is that the deformation away from the Minkowski
metric in the immediate neighborhood of a neutron star, measured
by $h_{00} \sim h_{ii} \sim 2GM/c^2 R$, where $R$ is the radius of
the star, is no longer a small quantity, as it
was in the solar system. In fact, while $h \sim 2GM/c^2 R$ is on
the order of $10^{-9}$ for the Earth and $10^{-6}$ for the Sun,
one finds that $h \sim 0.4$ for a typical neutron star ($M \simeq
1.4 \, M_{\odot}$, $R \sim 10$ km). One thus concludes that it is
no longer possible, as was the case in the solar system, to study
the structure and physics of neutron stars by using the
post-Newtonian approximation outlined above. One must consider the
exact form of Einstein's equations (\ref{rg9}), with all of their
non-linear structure. Because of this, we expect that observations
concerning neutron stars will allow us to confirm (or refute) the
theory of general relativity in its strongly non-linear regime. We
shall discuss such tests below in relation to observations of
binary pulsars.

\begin{figure}[h]
$$
\includegraphics[width=85mm]{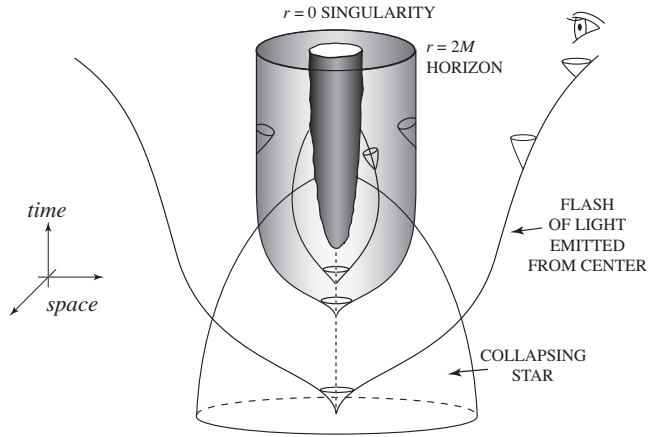}
$$
\caption{Schematic representation of the space-time for a black
hole created from the collapse of a spherical star. Each cone
represents the space-time history of a flash of light emitted from
a point at a particular instant. (Such a ``cone field'' is
obtained by taking the limit $\varepsilon^2 = 0$ from
Figure~\ref{fig2}, and keeping only the upper part, in other words
the part directed towards the future, of the double cones obtained
as limits of the hyperboloids of Figure~\ref{fig2}.) The interior
of the black hole is shaded, its outer boundary being the ``black
hole surface'' or ``horizon.'' The ``inner boundary'' (shown in
dark grey) of the interior region of the black hole is a
space-time singularity of the big-crunch type.}\label{fig3}
\end{figure}

A {\it black hole} is the result of a {\it continued}
collapse, meaning that it does not stop with the formation of an
ultra-dense star (such as a neutron star). (The physical concept
of a black hole was introduced by J.R.~Oppenheimer and H.~Snyder
in 1939. The global geometric structure of black holes was not
understood until some years later, thanks notably to the work of
R.~Penrose. For a historical review of the idea of black holes see
\cite{israel}.) It is a particular structure of curved space-time
characterized by the existence of a boundary (called the ``black
hole surface'' or ``horizon'') between an exterior region, from
which it is possible to emit signals to infinity, and an interior
region (of space-time), within which any emitted signal remains
trapped. See Figure~\ref{fig3}.

The cones shown in this figure are called ``light cones.'' They
are defined as the locus of points (infinitesimally close to $x$)
such that $ds^2 = 0$, with $dx^0 = c dt \geq 0$.
 Each represents the beginning of the space-time history of a flash
 of light emitted from a certain point in space-time. The cones whose vertices
 are located outside of the horizon (the shaded zone) will evolve by spreading
 out to infinity, thus representing the possibility for electromagnetic signals to reach infinity.

On the other hand, the cones whose vertices are located inside the
horizon (the grey zone) will evolve without ever succeeding in
escaping the grey zone. It is therefore impossible to emit an
electromagnetic signal that reaches infinity from the grey zone.
The horizon, namely the boundary between the shaded zone and the
unshaded zone, is itself the history of a particular flash of
light, emitted from the center of the star over the course of its
collapse, such that it asymptotically stabilizes as a space-time
cylinder. This space-time cylinder (the asymptotic horizon)
therefore represents the space-time history of a bubble of light
that, viewed locally, moves outward at the speed $c$, but which
globally ``runs in place.''  This remarkable behavior is a
striking illustration of the ``fluid'' character of space-time in
Einstein's theory.  Indeed, one can compare the preceding
situation with what may take place around the open drain of an
emptying sink: a wave may move along the water, away from the
hole, all the while running in place with respect to the sink
because of the falling motion of the water in the direction of the
drain.

Note that the temporal development of the interior region is
limited, terminating in a {\it singularity} (the dark gray
surface) where the curvature becomes infinite and where the
classical description of space and time loses its meaning. This
singularity is locally similar to the temporal inverse of a
cosmological singularity of the big bang type. This is called a
{\it big crunch}. It is a space-time frontier, beyond which
space-time ceases to exist. The appearance of singularities
associated with regions of strong gravitational fields is a
generic phenomenon in general relativity, as shown by  theorems
of R.~Penrose and S.W.~Hawking.

Black holes have some remarkable properties. First, a {\it
uniqueness} theorem (due to W.~Israel, B.~Carter, D.C.~Robinson,
G.~Bunting, and P.O.~Mazur) asserts that an isolated, stationary
black hole (in Einstein-Maxwell theory) is completely described by
three parameters: its mass $M$, its angular momentum $J$, and its
electric charge $Q$. The exact solution (called the Kerr-Newman
solution) of Einstein's equations (\ref{rg11}) describing a black
hole with parameters $M,J,Q$ is explicitly known. We shall here
content ourselves with writing the space-time geometry in the
simplest case of a black hole: the one in which $J=Q=0$ and the
black hole is described only by its mass (a solution discovered
by K.~Schwarzschild in January of 1916):
\begin{equation}
\label{rg17}
ds^2 = - \left( 1-\frac{2GM}{c^2 r} \right) c^2 \, dt^2 + \frac{dr^2}{1 -
\frac{2GM}{c^2 r}} + r^2 (d\theta^2 + \sin^2 \theta \, d\varphi^2) \, .
\end{equation}
We see that the purely temporal component of the metric, $g_{00} =
- (1 - 2GM / c^2 r)$, vanishes when the radial coordinate $r$
takes the value $r = r_H \equiv 2GM / c^2$. According to the
earlier equation (\ref{rg14}), it would therefore seem that light
emitted from an arbitrary point on the sphere $r_0 = r_H$, when it
is viewed by an observer located anywhere in the exterior (in $r
> r_H$), would experience an infinite reddening of its emission frequency ($
\nu / \nu_0 = 0$). In fact, the sphere $r_H = 2GM/c^2$ is the {\it
horizon} of the Schwarzschild black hole, and no particle (that is
capable of emitting light) can remain at rest when $r=r_H$ (nor, a
fortiori, when $r < r_H$). To study what happens at the horizon
($r = r_H$) or in the interior ($r < r_H$) of a Schwarzschild black
hole, one must use other space-time coordinates than the
coordinates $(t,r,\theta , \varphi)$ used in
Equation~(\ref{rg17}). The ``big crunch'' singularity in the
interior of a Schwarzschild black hole, in the coordinates of
(\ref{rg17}), is located at $r=0$ (which does not describe, as
one might believe, a point in space, but rather an instant in
time).

The space-time metric of a black hole space-time, such as
Equation~(\ref{rg17}) in the simple case $J=Q=0$, allows one to
study the influence of a black hole on particles and fields in its
neighborhood. One finds that a black hole is a gravitational
potential well that is so deep that any particle or wave that
penetrates the interior of the black hole (the region $r < r_H$)
will never be able to come out again, and that the total energy of the
particle or wave that ``falls'' into the black hole ends up augmenting
 the total mass-energy $M$ of the black hole. By studying such
black hole ``accretion'' processes with falling particles
(following R.~Penrose), D.~Christodoulou and R.~Ruffini showed
 that a black hole is not only a potential well, but also a
physical object possessing a significant
 {\it free energy} that it is possible, in principle, to extract.
 Such {\it black hole energetics} is encapsulated in the ``mass
 formula'' of Christodoulou and
Ruffini (in units where $c=1$)
\begin{equation}
\label{rg18}
M^2 = \left(M_{\rm irr} + \frac{Q^2}{4 \, GM_{\rm irr}} \right)^2 + \frac{J^2}{4
\, G^2M_{\rm irr}^2} \, ,
\end{equation}
where $M_{\rm irr}$ denotes the {\it irreducible mass} of the
black hole, a quantity {\it that can only grow, irreversibly}. One
deduces from (\ref{rg18}) that a rotating $(J \ne 0$) and/or
charged ($Q \ne 0$) black hole possesses a free energy $M - M_{\rm
irr} > 0$ that can, in principle, be extracted through processes
that reduce its angular momentum and/or its electric charge.  Such
black hole energy-extraction processes may lie at the origin of
certain ultra-energetic astrophysical phenomena (such as quasars
or gamma ray bursts). Let us note that, according to
Equation~(\ref{rg18}), (rotating or charged) black holes are the
largest reservoirs of free energy in the Universe: in fact, 29\%
of their mass energy can be stored in the form of rotational energy,
 and up to 50\% can be stored in the form of
electric energy. These percentages are much higher than the few
percent of nuclear binding energy that is at the origin of all the
light emitted by stars over their lifetimes. Even though there is
not, at present, irrefutable proof of the existence of black holes
in the universe, an entire range of very strong presumptive
evidence lends credence to their existence. In particular, more
than a dozen X-ray emitting binary systems in our galaxy are most
likely made up of a black hole and an ordinary star. Moreover, the
center of our galaxy seems to contain a very compact concentration
of mass $\sim 3 \times 10^6 M_{\odot}$
 that is probably a black hole. (For a review of the observational
 data leading to these conclusions see, for example, Section 7.6
of the recent book by N. Straumann \cite{livres}.)

The fact that a quantity associated with a black hole, here the
irreducible mass $M_{\rm irr}$, or, according to a more general
result due to S.W.~Hawking, the total area $A$ of the surface of a
black hole ($A = 16 \, \pi \, G^2 M_{\rm irr}^2$), can evolve only
by irreversibly growing is reminiscent of the second law of
thermodynamics. This result led J.D.~Bekenstein to interpret the
horizon area, $A$, as being proportional to the {\it entropy} of
the black hole. Such a thermodynamic interpretation is reinforced
by the study of the growth of $A$ under the influence of external
perturbations, a growth that one can in fact attribute to some
local dissipative properties of the black hole surface, notably a
surface viscosity and an electrical resistivity equal to 377 ohm
(as shown in work by T.~Damour and R.L.~Znajek). These
``thermodynamic'' interpretations of black hole properties are
based on simple analogies at the level of classical physics, but a
remarkable result by Hawking showed that they have real content at
the level of quantum physics. In 1974, Hawking discovered that the
presence of a horizon in a black hole space-time affected the
definition of a quantum particle, and caused a black hole to
continuously emit a flux of particles having the characteristic
spectrum (Planck spectrum) of thermal emission at the
temperature $T = 4 \, \hbar \, G \, \partial M / \partial A$,
where $\hbar$ is the reduced Planck constant. By using the general
thermodynamic relation connecting the temperature to the energy
$E=M$ and the entropy $S$, $T =
\partial M / \partial S$, we see from Hawking's result (in conformity
with Bekenstein's ideas) that a black hole possesses an {\it
entropy} $S$ equal (again with $c=1$) to
\begin{equation}
\label{rg19}
S = \frac{1}{4} \, \frac{A}{\hbar \, G} \, .
\end{equation}
The Bekenstein-Hawking formula (\ref{rg19}) suggests an
unexpected, and perhaps profound, connection between gravitation,
thermodynamics, and quantum theory. See Section~\ref{sec11} below.

\section{Binary Pulsars and Experimental Confirmations in the Regime
of Strong and Radiating Gravitational Fields}\label{sec9}

{\it Binary pulsars} are binary systems made up of a pulsar (a
rapidly spinning neutron star) and a very dense companion star
(either a neutron star or a white dwarf). The first system of this
type (called PSR B1913$ + $16) was discovered by  R.A.~Hulse and
J.H.~Taylor in 1974 \cite{hulse}. Today, a dozen are known. Some
of these (including the first-discovered PSR B1913$ + $16) have
revealed themselves to be remarkable probes of relativistic
gravitation and, in particular, of the regime of strong and/or
radiating gravitational fields. The reason for which a binary
pulsar allows for the probing of strong gravitational fields is
that, as we have already indicated above, the deformation of the
space-time geometry in the neighborhood of a neutron star is no
longer a small quantity, as it is in the solar system.  Rather, it
is on the order of unity: $h_{\mu\nu} \equiv g_{\mu\nu} -
\eta_{\mu\nu} \sim 2GM / c^2 R \sim 0.4$. (We note that this value
is only 2.5 times smaller than in the extreme case of a black
hole, for which $2GM / c^2 R = 1$.) Moreover, the fact that the
gravitational interaction propagates at the speed of light (as
indicated by the presence of the wave operator, $\Box = \Delta -
c^{-2}
\partial^2 / \partial t^2$ in (\ref{rg11})) between the pulsar and
its companion is found to play an observationally significant role
for certain binary pulsars.

Let us outline how the observational data from binary pulsars are
used to probe the regime of strong ($h_{\mu\nu}$ on the order of
unity) and/or radiative (effects propagating at the speed $c$)
gravitational fields. (For more details on the observational data
from binary pulsars and their use in probing relativistic
gravitation, see Michael Kramer's contribution to this Poincar\'e
seminar.) Essentially, a pulsar plays the role of an extremely
stable {\it clock}. Indeed, the ``pulsar phenomenon'' is due to
the rotation of a bundle of electromagnetic waves, created in the
neighborhood of the two magnetic poles of a strongly magnetized
neutron star (with a magnetic field on the order of $10^{12}$
Gauss, $10^{12}$ times the size of the terrestrial magnetic
field). Since the magnetic axis of a pulsar is not aligned with
its axis of rotation, the rapid rotation of the pulsar causes the
(inner) magnetosphere 
as a whole to rotate, and likewise the bundle of electromagnetic
waves created near the magnetic poles. The pulsar is therefore
analogous to a lighthouse that sweeps out space with two bundles
(one per pole) of electromagnetic waves. Just as for a lighthouse,
one does not see the pulsar from Earth except when the bundle
sweeps the Earth, thus causing a flash of electromagnetic noise
with each turn of the pulsar around itself (in some cases, one
even sees a secondary flash, due to emission from the second pole,
after each half-turn). One can then measure the time of arrival at
Earth of (the center of) each flash of electromagnetic noise. The
basic observational data of a pulsar are thus made up of a
regular, discrete sequence of the {\it arrival times} at Earth of
these flashes or ``pulses.''  This sequence is analogous to the
signal from a clock: tick, tick, tick, $\ldots$. Observationally, one
finds that some pulsars (and in particular those that belong to
binary systems)  thus define clocks of a stability comparable to
the best atomic clocks \cite{taylor}. In the case of a solitary
pulsar, the sequence of its arrival times is (in essence) a regular
``arithmetic sequence,'' $T_N = aN + b$, where $N$ is an integer
labelling the pulse considered, and where $a$ is equal to the
period of rotation of the pulsar around itself. In the case of a
binary pulsar, the sequence of arrival times is a much richer
signal, say $T_N = aN + b + \Delta_N$, where $\Delta_N$ measures
the deviation with respect to a regular arithmetic sequence. This
deviation (after the subtraction of effects not connected to the
orbital period of the pulsar) is due to a whole ensemble of
physical effects connected to the orbital motion of the pulsar
around its companion or, more precisely, around the center of mass
of the binary system. Some of these effects could be predicted by
a purely {\it Keplerian} description of the motion of the pulsar
in space, and are analogous to the ``R\oe mer effect'' that
allowed R\oe mer to determine, for the first time, the speed of
light from the arrival times at Earth of light signals coming from
Jupiter's satellites (the light signals coming from a body moving
in orbit are ``delayed'' by the time taken by light to cross this
orbit and arrive at Earth). Other effects can only be predicted
and calculated by using a {\it relativistic} description, either
of the orbital motion of the pulsar, or of the propagation of
electromagnetic signals between the pulsar and Earth. For example,
the following facts must be accounted for: (i) the ``pulsar
clock'' moves at a large speed (on the order of 300 km/s $\sim
10^{-3} c$) and is embedded in the varying gravitational potential of
the companion; (ii) the orbit of the pulsar is not a simple
Keplerian ellipse, but (in general relativity) a more complicated
orbit that traces out a ``rosette''  
 around the center of mass;
(iii) the propagation of electromagnetic signals between the
pulsar and Earth takes place in a space-time that is curved by
both the pulsar and its companion, which leads to particular
effects of relativistic delay; etc. Taking relativistic effects in
the theoretical description of arrival times for signals emitted
by binary pulsars into account thus leads one to write what is
called a {\it timing formula}. This timing formula (due
to T.~Damour and N.~Deruelle) in essence allows one to
parameterize the sequence of arrival times, $T_N = aN + b +
\Delta_N$, in other words to parameterize $\Delta_N$, as a
function of a set of ``phenomenological parameters'' that include
not only the so-called ``Keplerian'' parameters (such as the
orbital period $P$, the projection of the semi-major axis of the
pulsar's orbit along the line of sight $x_A = a_A \sin i$, and the
eccentricity $e$), but also the {\it post-Keplerian} parameters
associated with the relativistic effects mentioned above. For
example, effect (i) discussed above is parameterized by a quantity
denoted $\gamma_T$; effect (ii) by (among others) the quantities
$\dot\omega$, $\dot P$; effect (iii) by the quantities $r,s$; etc.

The way in which observations of binary pulsars allow one to test
relativistic theories of gravity is therefore the following. A
(least-squares) fit between the observational timing data,
$\Delta_N^{\rm obs}$, and the parameterized
theoretical timing formula, $\Delta_N^{\rm th} (P , x_A , e ; \gamma_T ,
\dot\omega , \dot P , r , s)$, allows for the determination of the
observational values of the Keplerian $(P^{\rm obs} , x_A^{\rm
obs} , e^{\rm obs})$ and post-Keplerian $(\gamma_T^{\rm obs} ,
\dot\omega^{\rm obs} , \dot P^{\rm obs} , r^{\rm obs} , s^{\rm
obs})$ parameters. The theory of general relativity predicts the
value of each post-Keplerian parameter as a function of the
Keplerian parameters and the two masses of the binary system (the
mass $m_A$ of the pulsar and the mass $m_B$ of the companion). For
example, the theoretical value predicted by general relativity for
the parameter $\gamma_T$ is $\gamma_T^{\rm GR} (m_A , m_B) =
en^{-1} (GMn/c^3)^{2/3} \, m_B (m_A + 2 \, m_B)/ M^2$, where $e$
is the eccentricity, $n = 2\pi / P$ the orbital frequency, and $M
\equiv m_A + m_B$. We thus see that, if one assumes that general
relativity is correct, the observational measurement of a
post-Keplerian parameter, for example $\gamma_T^{\rm obs}$,
determines a {\it curve} in the plane $(m_A , m_B)$ of the two
masses: $\gamma_T^{\rm GR} (m_A , m_B) = \gamma_T^{\rm obs}$, in
our example. The measurement of two post-Keplerian parameters thus
gives two curves in the $(m_A , m_B)$ plane and generically allows
one to determine the values of the two masses $m_A$ and $m_B$, by
considering the intersection of the two curves. We obtain a test
of general relativity as soon as one observationally measures
three or more post-Keplerian parameters: if the three (or more)
curves all intersect at one point in the plane of the two masses,
the theory of general relativity is confirmed, but if this is not
the case the theory is refuted. At present, four distinct binary
pulsars have allowed one to test general relativity. These four
``relativistic'' binary pulsars are: the first binary pulsar PSR
B1913$ + $16,
 the pulsar PSR B1534$ +$12 (discovered by A.~Wolszczan in 1991),
 and two recently discovered pulsars: PSR J1141$ - $6545 (discovered in 1999
 by V.M.~Kaspi et al., whose first timing results are due
 to M.~Bailes et al. in 2003), and PSR J0737$ - $3039 (discovered in 2003
by M.~Burgay et al., whose first timing results are due to
A.G.~Lyne et al. and M.~Kramer et al.). With the exception of PSR
J1141$ - $6545, whose companion is a white dwarf, the companions
of the pulsars are neutron stars. In the case of PSR J0737$ -
$3039 the companion turns out to also be a pulsar that is visible
from Earth.

\begin{figure}[h]
$$
\includegraphics[width=80mm]{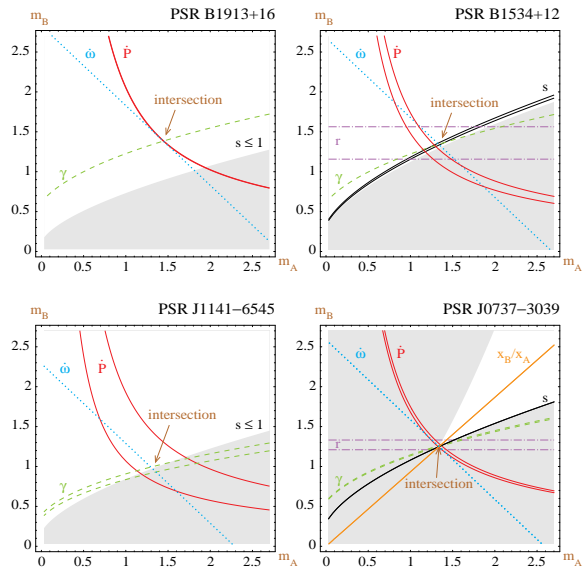}
$$
\caption{Tests of general relativity obtained from observations of
four binary pulsars. For each binary pulsar one has traced the
 ``curves,'' in the plane of the two masses ($m_A =$ mass of the pulsar, $m_B =$
mass of the companion), defined by equating the theoretical
expressions for the various post-Keplerian parameters, as
predicted by general relativity, to their observational value,
determined through a least-squares fit to the 
parameterized theoretical timing formula. Each ``curve'' is in fact a
``strip,'' whose thickness is given by the (one sigma) precision
with which the corresponding post-Keplerian parameter is measured.
For some parameters, these strips are too thin to be visible. The
grey zones would correspond to a sine for the angle of inclination
of the orbital plane with respect to the plane of the sky that is
greater than $1$, and are therefore physically
excluded.}\label{fig4}
\end{figure}

In the system PSR B1913$ + $16, {\it three} post-Keplerian
parameters have been measured $(\dot\omega , \gamma_T , \dot P)$,
which gives {\it one} test of the theory. In the system PSR J1141$
- $65, {\it three} post-Keplerian parameters have been measured
$(\dot\omega , \gamma_T , \dot P)$, which gives {\it one} test of
the theory. (The parameter $s$ is also measured through
scintillation phenomena, but the use of this measurement for
testing gravitation is more problematic.) In the system PSR B1534$
+ $12, {\it five} post-Keplerian parameters have been measured,
which gives {\it three} tests of the theory. In the system PSR
J0737$ - $3039,{\it six} post-Keplerian parameters,\footnote{In
the case of PSR J0737$ - $3039, one of the six measured parameters
is the ratio $x_A / x_B$ between a Keplerian parameter of the
pulsar and its analog for the companion, which turns out to also
be a pulsar.} which gives {\it four} tests of the theory. It is
remarkable that all of these tests have confirmed general
relativity. See Figure~\ref{fig4} and, for references and details,
\cite{tests,taylor,stairs,gef}, as well as the contribution by
Michael Kramer.

Note that, in Figure~\ref{fig4}, some post-Keplerian parameters
are measured with such great precision that they in fact define
very thin curves in the $m_A , m_B$ plane. On the other hand, some
of them are only measured with a rough fractional precision and
thus define ``thick curves,'' or ``strips'' in the plane of the
masses (see, for example, the strips associated with $\dot P$, $r$
and $s$ in the case of PSR B1534$ + $12). In any case, the theory
is confirmed when all of the strips (thick or thin) have a
non-empty common intersection. (One should also note that the
strips represented in Figure~\ref{fig4} only use the ``one sigma''
error bars, in other words a 68\% level of confidence.
Therefore, the fact that the $\dot P$ strip for PSR B1534$ + $12
is a little bit disjoint from the intersection of the other strips
is not significant: a ``two sigma'' figure would show excellent
agreement between observation and general relativity.)

In view of the arguments presented above, all of the tests shown
in Figure~\ref{fig4} confirm the validity of general relativity in
the regime of strong gravitational fields ($h_{\mu\nu} \sim 1$).
Moreover, the four tests that use measurements of the parameter
$\dot P$ (in the four corresponding systems) are direct
experimental confirmations of the fact that the gravitational
interaction propagates at the speed $c$ between the companion and
the pulsar. In fact, $\dot P$ denotes the long-term variation
$\langle dP / dt \rangle$ of the orbital period. Detailed
theoretical calculations of the motion of two gravitationally
condensed objects in general relativity, that take into account
the effects connected to the propagation of the gravitational
interaction at finite speed\cite{motion}, have shown that one of
the observable effects of this propagation is a long-term decrease
in the orbital period given by the formula
$$
\dot P^{\rm GR} (m_A , m_B) = - \frac{192 \, \pi}{5} \, \frac{1 + \frac{73}{24}
\, e^2 + \frac{37}{96} \, e^4}{(1-e^2)^{7/2}} \left( \frac{GM n}{c^3}
\right)^{5/3} \, \frac{m_A \, m_B}{M^2} \, .
$$
The direct physical origin of this decrease in the orbital period
lies in the modification, produced by general relativity, of the
usual Newtonian law of gravitational attraction between two
bodies, $F_{\rm Newton} = G \, m_A \, m_B / r_{AB}^2$. In place of
such a simple law, general relativity predicts a more complicated
force law that can be expanded in the symbolic form
\begin{equation}
\label{rgn19}
F_{\rm Einstein} = \frac{G \, m_A \, m_B}{r_{AB}^2} \left( 1 + \frac{v^2}{c^2} +
\frac{v^4}{c^4} + \frac{v^5}{c^5} + \frac{v^6}{c^6} + \frac{v^7}{c^7} + \cdots
\right) \, ,
\end{equation}
where, for example, ``$v^2 / c^2$'' represents a whole set of
terms of order $v_A^2 / c^2$, $v_B^2 / c^2$, $v_A \, v_B / c^2$,
or even $G \, m_A / c^2 \, r$ or $G \, m_B / c^2 \, r$. Here $v_A$
denotes the speed of body $A$, $v_B$ that of body $B$, and
$r_{AB}$ the distance between the two bodies. The term of order
$v^5 / c^5$ in Equation~(\ref{rgn19}) is particularly important.
This term is a direct consequence of the finite-speed propagation
of the gravitational interaction between $A$ and $B$, and its
calculation shows that it contains a component that is opposed to
the relative speed $\vv_A - \vv_B$ of the two bodies and that,
consequently, slows down the orbital motion of each body, causing
it to evolve towards an orbit that lies closer to its companion
(and therefore has a shorter orbital period). This ``braking''
term (which is correlated with the emission of gravitational
waves), $\delta F_{\rm Einstein} \sim v^5 / c^5 \, F_{\rm
Newton}$, leads to a long-term decrease in the orbital period
$\dot P^{\rm GR} \sim - (v/c)^5 \sim -10^{-12}$ that is very
small, but whose reality has been verified with a fractional
precision of order $10^{-3}$ in PSR B1913$ + $16 and of order 20\%
in PSR B1534$ + $12 and PSR J1141$ - $6545
\cite{tests,taylor,gef}.

To conclude this brief outline of the tests of relativistic
gravitation by binary pulsars, let us note that there is an
analog, for the regime of strong gravitational fields, of the
formalism of parametrization for possible deviations from general
relativity mentioned in Section~\ref{sec6} in the framework of
weak gravitational fields (using the post-Newtonian parameters
$\gamma - 1$ and $\beta - 1$). This analog is obtained by
considering a two-parameter family of relativistic theories of
gravitation, assuming that the gravitational interaction is
propagated not only by a tensor field $g_{\mu\nu}$ but also by a
scalar field $\varphi$. Such a class of tensor-scalar theories of
gravitation allows for a description of possible deviations in
both the solar system and in binary pulsars. It also allows one to
explicitly demonstrate that binary pulsars indeed test the effects
of strong fields that go beyond the tests of the weak fields of
the solar system by exhibiting classes of theories that are
compatible with all of the observations in the solar system but
that are incompatible with the observations of binary pulsars, see
\cite{tests,gef}.

\section{Gravitational Waves: Propagation, Genera\-tion, and
Detection}\label{sec10}

As soon as he had finished constructing the theory of general
relativity, Einstein realized that it implied the existence of
waves of geometric deformations of space-time, or ``gravitational
waves'' \cite{E16,oeuvres}. Mathematically, these waves are
analogs (with the replacement $A_{\mu} \to h_{\mu\nu}$) of
electromagnetic waves, but conceptually they signify something
remarkable: they exemplify, in the purest possible way, the
``elastic'' nature of space-time in general relativity. Before
Einstein space-time was a rigid structure, given a priori, which
was not influenced by the material content of the Universe. After
Einstein, a distribution of matter (or more generally of
mass-energy) that changes over the course of time, let us say for
concreteness a binary system of two neutron stars or two black
holes, will not only deform the chrono-geometry of the space-time
in its immediate neighborhood, but this deformation will propagate
in every possible direction away from the system considered, and
will travel out to infinity in the form of a wave whose
oscillations will reflect the temporal variations of the matter
distribution. We therefore see that the study of these
gravitational waves poses three separate problems: that of
generation, that of propagation, and, finally, that of detection
of such gravitational radiation. These three problems are at
present being actively studied, since it is hoped that we will
soon detect gravitational waves, and thus will be able to obtain
new information about the Universe \cite{thorne}. We shall here
content ourselves with an elementary introduction to this field of
research. For a more detailed introduction to the detection of
gravitational waves see the contribution by Jean-Yves Vinet to
this Poincar\'e seminar.

Let us first consider the simplest case of very weak gravitational
waves, outside of their material sources. The geometry of such a
space-time can be written, as in Section~\ref{sec6}, as
$g_{\mu\nu} (x) = \eta_{\mu\nu} + h_{\mu\nu} (x)$, where
$h_{\mu\nu} \ll 1$. At first order in $h$, and outside of the
source (namely in the domain where $T_{\mu\nu} (x) = 0$), the
perturbation of the geometry, $h_{\mu\nu} (x)$, satisfies a
homogeneous equation obtained by replacing the right-hand side of
Equation~(\ref{rg11}) with zero. It can be shown that one can
simplify this equation through a suitable choice of coordinate
system. In a {\it transverse traceless} (TT) coordinate system the
only non-zero components of a general gravitational wave are the
spatial components $h_{ij}^{\rm TT}$, $i,j = 1,2,3$ (in other
words $h_{00}^{\rm TT} = 0 = h_{0i}^{\rm TT}$), and these
components satisfy
\begin{equation}
\label{rg20}
\Box \, h_{ij}^{\rm TT} = 0 \, , \ \partial_j \, h_{ij}^{\rm TT} = 0 \, , \
h_{jj}^{\rm TT} = 0 \, .
\end{equation}
The first equation in (\ref{rg20}), where the wave operator $\Box
= \Delta - c^{-2} \, \partial_t^2$ appears, shows that
gravitational waves (like electromagnetic waves) propagate at the
speed $c$. If we consider for simplicity a monochromatic plane
wave ($h_{ij}^{\rm TT} = \zeta_{ij} \, \exp (i \, \vk \cdot \vx -
i \, \omega \, t) \, +$ complex conjugate, with $\omega = c \,
\vert \vk \vert$), the second equation in (\ref{rg20}) shows that
the (complex) tensor $\zeta_{ij}$ measuring the polarization of a
gravitational wave only has non-zero components in the plane
orthogonal to the wave's direction of propagation: $\zeta_{ij} \,
k^j = 0$. Finally, the third equation in (\ref{rg20}) shows that
the polarization tensor $\zeta_{ij}$ has vanishing trace:
$\zeta_{jj} = 0$. More concretely, this means that if a
gravitational wave propagates in the $z$-direction, its
polarization is described by a $2 \times 2$ matrix,
$\begin{pmatrix} \zeta_{xx} &\zeta_{xy}
\\ \zeta_{yx} &\zeta_{yy}
\end{pmatrix}$, which is symmetric and traceless. Such a polarization matrix
therefore only contains two independent (complex) components:
$\zeta_+ \equiv \zeta_{xx} = - \zeta_{yy}$, and $\zeta_{\times}
\equiv \zeta_{xy} = \zeta_{yx}$. This is the same number of
independent (complex) components that an electromagnetic wave has.
Indeed, in a transverse gauge, an electromagnetic wave only has
spatial components $A_i^T$ that satisfy
\begin{equation}
\label{rg21}
\Box \, A_i^T = 0 \, , \ \partial_j \, A_j^T = 0 \, .
\end{equation}
As in the case above, the first equation (\ref{rg21}) means that
an electromagnetic wave propagates at the speed $c$, and the
second equation shows that a monochromatic plane electromagnetic
wave ($A_i^T = \zeta_i \, \exp (i \, \vk \cdot \vx - i \, \omega
\, t) +$ c.c., $\omega = c \, \vert \vk \vert$) is described by a
(complex) polarization vector $\zeta_i$ that is orthogonal to the direction
of propagation: $\zeta_j \, k^j = 0$. For a wave propagating in
the $z$-direction such a vector only has two independent (complex)
components, $\zeta_x$ and $\zeta_y$. It is indeed the same number
of components that a gravitational wave has, but we see that the
two quantities measuring the polarization of a gravitational wave,
$\zeta_+ = \zeta_{xx} = - \zeta_{yy}$, $\zeta_{\times} =
\zeta_{xy} = \zeta_{yx}$ are mathematically quite different from
the two quantities $\zeta_x , \zeta_y$ measuring the polarization
of an electromagnetic wave. However, see Section~\ref{sec11}
below. 

We have here discussed the propagation of a gravitational
wave in a background space-time described by the Minkowski metric
$\eta_{\mu\nu}$. One can also consider the propagation of a wave
in a curved background space-time, namely by studying solutions of
Einstein's equations (\ref{rg9}) of the form $g_{\mu\nu} (x) =
g_{\mu\nu}^B (x) + h_{\mu\nu} (x)$ where $h_{\mu\nu}$ is not only
small, but varies on temporal and spatial scales much shorter than
those of the background metric $g_{\mu\nu}^B (x)$. Such a study is
necessary, for example, for understanding the propagation of
gravitational waves in the cosmological Universe.

The problem of {\it generation} consists in searching for the
connection between the tensorial amplitude $h_{ij}^{\rm TT}$ of
the gravitational radiation in the radiation zone and the motion
and structure of the source. If one considers the simplest case of
a source that is sufficiently diffuse that it only creates waves
that are everywhere weak ($g_{\mu\nu} - \eta_{\mu\nu} = h_{\mu\nu}
\ll 1$), one can use the linearized approximation to Einstein's
equations (\ref{rg9}), namely Equations (\ref{rg11}). One can
solve Equations (\ref{rg11}) by the same technique that is used to
solve Maxwell's equations (\ref{rg12}): one fixes the coordinate
system by imposing $\partial^{\alpha} \, h_{\alpha\mu} -
\frac{1}{2} \,
\partial_{\mu} \, h_{\alpha}^{\alpha} = 0$ (analogous to the Lorentz
gauge condition $\partial^{\alpha} \, A_{\alpha} = 0$), then one
inverts the wave operator by using retarded potentials. Finally,
one must study the asymptotic form, at infinity, of the emitted
wave, and write it in the reduced form of a transverse and
traceless amplitude $h_{ij}^{\rm TT}$ satisfying
Equations~(\ref{rg20}) (analogous to a transverse electromagnetic
wave $A_i^T$ satisfying (\ref{rg21})). One then finds that, just
as charge conservation implies that there is no monopole type
electro-magnetic radiation, but only dipole or higher orders of polarity,
the conservation of energy-momentum implies the absence of
monopole {\it and} dipole gravitational radiation. For a slowly
varying source $(v/c \ll 1$), the dominant gravitational radiation
is of {\it quadrupole} type. It is given, in the radiation zone,
by an expression of the form
\begin{equation}
\label{rg22}
h_{ij}^{TT} (t,r,\vn) \simeq \frac{2G}{c^4 \, r} \, \frac{\partial^2}{\partial
t^2} \, [ I_{ij} (t-r/c)]^{\rm TT} \, .
\end{equation}
Here $r$ denotes the distance to the center of mass of the
source, $I_{ij} (t) \equiv \int d^3 x \, c^{-2}$ $T^{00} (t,\vx)
\left(x^i x^j - \frac{1}{3} \, \vx^2 \delta^{ij} \right)$ is the
quadrupole moment of the mass-energy distribution, and the upper
index TT denotes an algebraic projection operation for the
quadrupole tensor $I_{ij}$ (which is a $3 \times 3$ matrix) that
only retains the part orthogonal to the local direction of wave
propagation $n^i \equiv x^i / r$ with vanishing trace
($I_{ij}^{\rm TT}$ is therefore locally a (real) $2 \times 2$ symmetric,
traceless matrix of the same type as $\zeta_{ij}$ above). Formula
(\ref{rg22}) (which was in essence obtained by Einstein in 1918
\cite{E16}) is only the first approximation to an expansion in
powers of $v/c$, where $v$ designates an internal speed
characteristic of the source. The prospect of soon being able to
detect gravitational waves has motivated theorists to improve
Formula (\ref{rg22}): (i) by describing the terms of higher order
in $v/c$, up to a very high order, and (ii) by using new
approximation methods that allow one to treat sources containing
regions of strong gravitational fields (such as, for example, a
binary system of two black holes or two neutron stars). See below
for the most recent results.

Finally, the problem of {\it detection}, of which the pioneer was
Joseph Weber in the 1960s, is at present giving rise to very
active experimental research. The principle behind any detector is
that a gravitational wave of amplitude $h_{ij}^{\rm TT}$ induces a
change in the distance $L$ between two bodies on the order of
$\delta L \sim hL$ during its passage. One way of seeing this is
to consider the action of a wave $h_{ij}^{\rm TT}$ on two free
particles, at rest before the arrival of the wave at the positions
$x_1^i$ and $x_2^i$ respectively. As we have seen, each particle,
in the presence of the wave, will follow a geodesic motion in the
geometry $g_{\mu\nu} = \eta_{\mu\nu} + h_{\mu\nu}$ (with $h_{00} =
h_{0i} = 0$ and $h_{ij} = h_{ij}^{\rm TT}$). By writing out the
geodesic equation, Equation (\ref{rg7}), one finds that it simply
reduces (at first order in $h$) to $d^2 x^i / ds^2 = 0$.
Therefore, particles that are initially at rest ($x^i =$ const.)
remain at rest in a transverse and traceless system of
coordinates! This does not however mean that the gravitational
wave has no observable effect. In fact, since the spatial geometry
is perturbed by the passage of the wave, $g_{ij} (t,\vx) =
\delta_{ij} + h_{ij}^{\rm TT} (t,\vx)$, one finds that the
physical distance between the two particles $x_1^i$, $x_2^i$
(which is observable, for example, by measuring the time taken for
light to make a round trip between the two particles) varies,
during the passage of the wave, according to $L^2 = (\delta_{ij} +
h_{ij}^{\rm TT}) (x_2^i - x_1^i) (x_2^j - x_1^j)$.

The problem of detecting a gravitational wave thus leads to the
problem of detecting a small relative displacement $\delta L / L
\sim h$. By using Formula (\ref{rg22}), one finds that the order
of magnitude of $h$, for known or hoped for astrophysical sources
(for example,a very close system of two neutron stars or two black
holes), situated at distances such that one may hope to see
several events per year ($r \gtrsim 600$ million light-years), is
in fact extremely small: $h \lesssim 10^{-22}$ for signals whose
characteristic frequency is around 100 Hertz. Several types of
detectors have been developed since the pioneering work of
J.~Weber \cite{thorne}. At present, the detectors that should
succeed in the near future at detecting amplitudes $h \sim \delta
L / L \sim 10^{-22}$ are large interferometers, of the Michelson
or  Fabry-P\'erot type, having arms that are many kilometers in
length into which a very powerful monochromatic laser beam is
injected. Such terrestrial interferometric detectors presently
exist in the U.S.A. (the LIGO detectors \cite{ligo}), in Europe
(the VIRGO \cite{virgo} and GEO 600 \cite{geo} detectors) and
elsewhere (such as the TAMA detector in Japan). Moreover, the
international space project LISA \cite{lisa}, made up of an
interferometer between satellites that are several million
kilometers apart, should allow one to detect low frequency ($\sim$
one hundredth or one thousandth of a Hertz) gravitational waves in
a dozen years or so. This collection of gravitational wave
detectors promises to bring invaluable information for astronomy
by opening a new ``window'' on the Universe that is much more
transparent than the various electromagnetic (or neutrino) windows
that have so greatly expanded our knowledge of the Universe in the
twentieth century.

\begin{figure}[h]
$$
\includegraphics[width=80mm]{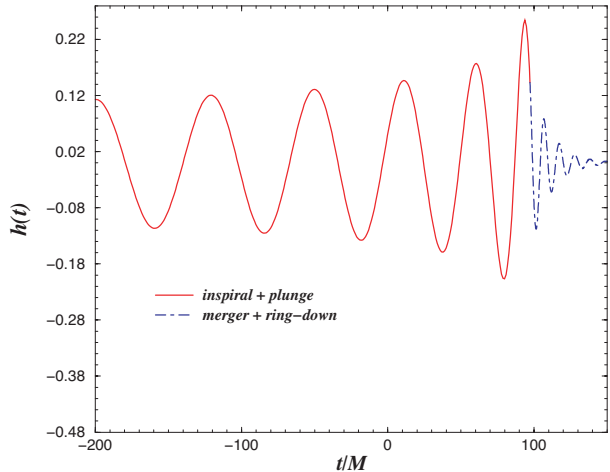}
$$
\caption{The gravitational amplitude $h(t)$ emitted during the
final stages of evolution of a system of two equal-mass black
holes. The beginning of the signal (the left side of the figure),
which is sinusoidal, corresponds to an inspiral motion 
 of two separate black holes (with decreasing distance); 
 the middle corresponds to a
rapid ``inspiralling plunge'' of the two black holes towards each
other; the end (at right) corresponds to the oscillations of the
final, rotating black hole formed from the merger of the two
initial black holes.}\label{fig5}
\end{figure}

The extreme smallness of the expected gravitational signals has
led a number of experimentalists to contribute, over many years, a
wealth of ingenuity and know-how in order to develop technology
that is sufficiently precise and trustworthy (see
\cite{ligo,virgo,geo,lisa}). To conclude, let us also mention how
much concerted theoretical effort has been made, both in
calculating the general relativistic predictions for gravitational
waves emitted by certain sources, and in developing methods
adapted to the extraction of the gravitational signal from the
background noise in the detectors. For example, one of the most
promising sources for terrestrial detectors is the wave train for
gravitational waves emitted by a system of two black holes, and in
particular the final (most intense) portion of this wave train,
which is emitted during the last few orbits of the system and the
final coalescence of the two black holes into a single, more
massive black hole. We have seen above (see Section~\ref{sec9})
that the finite speed of propagation of the gravitational
interaction between the two bodies of a binary system gives rise
to a progressive acceleration of the orbital frequency, connected
to the progressive approach of the two bodies towards each other.
Here we are speaking of the final stages in such a process, where
the two bodies are so close that they orbit around each other in a
spiral pattern that accelerates until they attain (for the final
``stable'' orbits) speeds that become comparable to the speed of
light, all the while remaining slightly slower. In order to be
able to determine, with a precision that is acceptable for the
needs of detection,  the dynamics of such a binary black hole
system in such a situation, as well as the gravitational amplitude
$h_{ij}^{\rm TT}$that it emits, it was necessary to develop a
whole ensemble of analytic techniques to a very high level of
precision. For example, it was necessary to calculate the
expansion (\ref{rgn19}) of the force determining the motion of the
two bodies to a very high order and also to calculate the
amplitude $h_{ij}^{\rm TT}$ of the gravitational radiation emitted
to infinity with a precision going well beyond the quadrupole
approximation (\ref{rg22}). These calculations are comparable in
complexity to high-order calculations in quantum field theory.
Some of the techniques developed for quantum field theory indeed
proved to be extremely useful for these calculations in the
(classical) theory of general relativity (such as certain
resummation methods and the mathematical use of analytic
continuation in the number of space-time dimensions). For an
entryway into the literature of these modern analytic methods, see
\cite{BDEI04}, and for an early example of a result obtained by such
methods of direct interest for the physics of detection see
Figure~\ref{fig5} \cite{BD}, which shows a component of the
gravitational amplitude $h_{ij}^{\rm TT} (t)$ emitted during the
final stages of evolution of a system of two black holes of equal
mass. The first oscillations shown in Figure~\ref{fig5} are
emitted during the last quasi-circular orbits (accelerated motion
in a spiral of decreasing radius). The middle part of the signal
corresponds to a phase where, having moved past the last stable
orbit, the two black holes ``fall'' toward each other while
spiraling rapidly. In fact, contrarily to Newton's theory, which
predicts that two condensed bodies would be able to orbit around
each other with an orbit of arbitrarily small radius (basically up
until the point that the two bodies touch), Einstein's theory
predicts a modified law for the force between the two bodies,
Equation~(\ref{rgn19}), whose analysis shows that it is so
attractive that it no longer allows for stable circular orbits
when the distance between the two bodies becomes smaller than
around $6 \, G (m_A + m_B) / c^2$. In the case of two black holes,
this distance is sufficiently larger than the black hole ``radii''
 ($2 \, G \, m_A /
c^2$ and $2 \, G \, m_B / c^2$) that one is still able to
analytically treat the beginning of the ``spiralling plunge'' 
of the two black holes towards each other. The final oscillations
in Figure~\ref{fig5} are emitted by the rotating (and initially
highly deformed) black hole formed from the merger of the two
initial, separate black holes.

Up until quite recently the analytic predictions illustrated in
Figure~\ref{fig5} concerning the gravitational signal $h(t)$
emitted by the spiralling plunge and merger of two black holes
remained conjectural, since they could be compared to neither
other theoretical predictions nor to observational data. Recently,
worldwide efforts made over three decades to attack the
problem of the coalescence of two black holes by {\it numerically}
solving Einstein's equations (\ref{rg9}) have spectacularly begun to bear fruit.
Several groups have been able to numerically calculate the signal
$h(t)$ emitted during the final orbits and merger of two black
holes \cite{NR}. In essence, there is good agreement between the
analytical and numerical predictions. In order to be able to
detect the gravitational waves emitted by the coalescence of two
black holes, it will most likely be necessary to properly combine
the information on the structure of the signal $h(t)$ obtained by
the two types of methods, which are in fact complementary.

\section{General Relativity and Quantum Theory: From Supergravity to String Theory}\label{sec11}

Up until now, we have discussed the {\it classical} theory of
general relativity, neglecting any quantum effects.  What becomes
of the theory in the quantum regime? This apparently innocent
question in fact opens up vast new prospects that are still under
construction.  We will do nothing more here than to touch upon the
subject, by pointing out to the reader some of the paths along
which contemporary physics has been led by the challenge of
unifying general relativity and quantum theory. For a more
complete introduction to the various possibilities ``beyond''
general relativity suggested within the framework of {\it string
theory} (which is still under construction) one should consult the
contribution of Ignatios Antoniadis to this Poincar\'e Seminar.

Let us recall that, from the very beginning of the
quasi-definitive formulation of quantum theory (1925--1930), the
creators of quantum mechanics (Born, Heisenberg, Jordan; Dirac;
Pauli; etc.) showed how to ``quantize'' not only 
systems with several particles (such as an atom), but also {\it
fields}, continuous dynamical systems whose classical description
implies a continuous distribution of energy and momentum in space.
In particular, they showed how to {\it quantize} (or in other
words how to formulate within a framework compatible with quantum
theory) the electromagnetic field $A_{\mu}$, which, as we have
recalled above, satisfies the Maxwell equations (\ref{rg12}) at
the classical level. They nevertheless ran into difficulty due to
the following fact. In quantum theory, the physics of a system's
evolution is essentially contained in the {\it transition
amplitudes} $A (f,i)$ between an initial state labelled by $i$ and
a final state labelled by $f$. These amplitudes $A(f,i)$ are
complex numbers. They satisfy a ``transitivity'' property of the
type
\begin{equation}
\label{rg23}
A(f,i) = \sum_n A(f,n) \, A(n,i) \, ,
\end{equation}
which contains a sum over all possible intermediate states,
labelled by $n$ (with this sum becoming an integral when there is
a continuum of intermediate possible states). R.~Feynman used
Equation~(\ref{rg23}) as a point of departure for a new
formulation of quantum theory, by interpreting it as an analog of
{\it Huygens' Principle}: if one thinks of $A(f,i)$ as the
amplitude, ``at the point $f$,'' of a ``wave'' emitted ``from the
point $i$,'' Equation~(\ref{rg23}) states that this amplitude can
be calculated by considering the ``wave'' emitted from $i$ as
passing through all possible intermediate ``points'' $n$
($A(n,i)$), while reemitting  ``wavelets'' starting from these
intermediate points ($A(f,n)$), which then superpose to form the
total wave arriving at the ``final point $f$.''

Property (\ref{rg23}) does not pose any problem in the quantum
mechanics of discrete systems (particle systems). It simply shows
that the amplitude $A(f,i)$ behaves like a wave, and therefore
must satisfy a ``wave equation'' (which is indeed the case for the
Schr\"odinger equation describing the dependence of $A(f,i)$ on
the parameters determining the final configuration $f$). On the
other hand, Property (\ref{rg23}) poses formidable problems when
one applies it to the quantization of continuous dynamical systems
(fields). In fact, for such systems the ``space'' of intermediate
possible states is infinitely larger than in the case of the
mechanics of discrete systems.  Roughly speaking, the intermediate
possible states for a field can be described as containing  $\ell
= 1,2,3,\ldots$ quantum excitations of the field, with each
quantum excitation (or pair of ``virtual particles'') being
described essentially by a plane wave, $\zeta \exp (i \, k_{\mu}
\, x^{\mu})$, where $\zeta$ measures the polarization of these
virtual particles and $k^{\mu} = \eta^{\mu\nu} \, k_{\nu}$, with
$k^0 = \omega$ and $k^i = \vk$, their angular frequency and wave
vector, or (using the Planck-Einstein-de Broglie relations $E =
\hbar \, \omega$, $\vp = \hbar \, \vk$) their energy-momentum
$p^{\mu} = \hbar \, k^{\mu}$. The quantum theory shows (basically
because of the uncertainty principle) that the four-frequencies
(and four-momenta) $p^{\mu} = \hbar \, k^{\mu}$ of the
intermediate states cannot be constrained to satisfy the classical
equation $\eta_{\mu\nu} \, p^{\mu} \, p^{\nu} = -m^2$ (or in other
words $E^2 = \vp^2 + m^2$~; we use $c=1$ in this section). As a
consequence, the sum over intermediate states for a quantum field
theory has the following properties (among others): (i) when $\ell
= 1$ (an intermediate state containing only one pair of virtual
particles, called a {\it one-loop contribution}), there is an
integral over a four-momentum $p^{\mu}$, $\int d^4 p = \int d E
\int d \vp$; (ii) when $\ell = 2$ (two pairs of virtual particles;
a {\it two-loop contribution}), there is an integral over two
four-momenta $p_1^{\mu}$, $p_2^{\mu}$, $\int d^4 p_1 \, d^4 p_2$;
etc. The delicate point comes from the fact that the
energy-momentum of an intermediate state can take arbitrarily high
values. This possibility is directly connected (through a Fourier
transform) to the fact that a field possesses an infinite number
of degrees of freedom, corresponding to configurations that vary
over arbitrarily small time and length scales.

The problems posed by the necessity of integrating over the
infinite domain of four-momenta of intermediate virtual particles
(or in other words of accounting for the fact that field
configurations can vary over arbitrarily small scales) appeared in
the 1930s when the quantum theory of the electromagnetic field
$A_{\mu}$ (called quantum electrodynamics, or QED) was studied in
detail. These problems imposed themselves in the following form:
when one calculates the transition amplitude for given initial and
final states (for example the collision of two light quanta,
with two photons entering and two photons leaving) by using
(\ref{rg23}), one finds a result given in the form of a {\it
divergent integral}, because of the integral (in the one-loop
approximation, $\ell = 1$) over the arbitrarily large
energy-momentum describing virtual electron-positron pairs
appearing as possible intermediate states. Little by little,
theoretical physicists understood that the types of divergent
integrals appearing in QED were relatively benign and, after the
second world war, they developed a method ({\it renormalization
theory}) that allowed one to unambiguously isolate the infinite
part of these integrals, and to subtract them by expressing the
amplitudes $A(f,i)$ solely as a function of observable quantities
\cite{weinberg} (work by J.~Schwinger, R.~Feynman, F.~Dyson etc.).

The preceding work led to the development of consistent quantum
theories not only for the electromagnetic field $A_{\mu}$ (QED),
but also for generalizations of electromagnetism (Yang-Mills
theory or non-abelian gauge theory) that turned out to provide
excellent descriptions of the new interactions between elementary
particles discovered in the twentieth century (the electroweak
theory, partially unifying electromagnetism and weak nuclear
interactions, and quantum chromodynamics, describing the strong
nuclear interactions). All of these theories give rise to only
relatively benign divergences that can be ``renormalized'' and
thus allowed one to compute amplitudes $A(f,i)$ corresponding
to observable physical processes \cite{weinberg} (notably, work by
G.~'t~Hooft and M.~Veltman).

What happens when we use (\ref{rg23}) to construct a
``perturbative'' quantum theory of general relativity (namely one
obtained by expanding in the number $\ell$ of virtual particle
pairs appearing in the intermediate states)? The answer is that
the integrals over the four-momenta of intermediate virtual
particles are not at all of the benign type that allowed them to
be renormalized in the simpler case of electromagnetism. The
source of this difference is not accidental, but is rather
connected with the basic physics of relativistic gravitation.
Indeed, as we have mentioned, the virtual particles have
arbitrarily large energies $E$. Because of the basic relations
that led Einstein to develop general relativity, namely $E = m_i$
and $m_i = m_g$, one deduces that these virtual particles
correspond to arbitrarily large gravitational masses $m_g$. They
will therefore end up creating intense gravitational effects that
become more and more intense as the number $\ell$ of virtual
particle pairs grows. These gravitational interactions that grow
without limit with energy and momentum correspond (by Fourier
transform) to field configurations concentrated in arbitrarily
small space and time scales. One way of seeing why the quantum
gravitational field creates much more violent problems than the
quantum electromagnetic field is, quite simply, to go back to
dimensional analysis. Simple considerations in fact show that the
relative (non-dimensional) one-loop amplitude $A_1$ must be
proportional to the product $\hbar \, G$ and must contain an
integral $\int d^4 k$. However, in 1900 Planck had noticed that
(in units where $c=1$) the dimensions of $\hbar$ and $G$ were such
that the product $\hbar \, G$ had the dimensions of length (or
time) squared:
\begin{equation}
\label{rg24}
\ell_P \equiv \sqrt{\frac{\hbar \, G}{c^3}} \simeq 1.6 \times 10^{-33} \, {\rm
cm} , \ t_P \equiv \sqrt{\frac{\hbar \, G}{c^5}} \simeq 5.4 \times 10^{-44} \,
{\rm s} \, .
\end{equation}
One thus deduces that the integral $\int d^4 k \, f(k)$ must have
the dimensions of a squared frequency, and therefore that $A_1$
must (when $k \to \infty$) be of the type, $A_1 \sim \hbar \, G
\int d^4 k / k^2$. Such an integral diverges quadratically with
the upper limit $\Lambda$ of the integral (the cutoff frequency,
such that $\vert k \vert \leq \Lambda$), so that $A_1
\sim \hbar \, G \, \Lambda^2 \sim t_P^2 \, \Lambda^2$. The
extension of this dimensional analysis to the intermediate states
with several loops ($\ell > 1$) causes even more severe polynomial
divergences to appear, of a type such that the power of $\Lambda$
that appears grows without limit with $\ell$.

In summary, the essential physical characteristics of gravitation
($E = m_i = m_g$ and the dimension of Newton's constant $G$) imply
the impossibility of generalizing to the gravitational case
the methods that allowed a
satisfactory quantum treatment of the other interactions
(electromagnetic, weak, and strong).
Several paths have been explored to get out of this impasse. Some
researchers tried to quantize general relativity
non-perturbatively, without using an expansion in intermediate
states (\ref{rg23}) (work by A.~Ashtekar, L.~Smolin, and others).
others have tried to generalize general relativity by adding a
fermionic field to Einstein's (bosonic) gravitational field
$g_{\mu\nu} (x)$, the gravitino field $\psi_{\mu} (x)$. It is
indeed remarkable that it is possible to define a theory, known as
{\it supergravity}, that generalizes the geometric invariance of
general relativity in a profound way. After the 1974 discovery (by
J.~Wess and B.~Zumino) of a possible new global symmetry for
interacting bosonic and fermionic fields, {\it supersymmetry}
(which is a sort of global rotation transforming bosons to
fermions and vice versa), D.Z.~Freedman, P.~van~Nieuwenhuizen, and
S.~Ferrara; and S.~Deser and B.~Zumino; showed that one could
generalize global supersymmetry to a {\it local supersymmetry},
meaning that it varies from point to point in space-time. Local
supersymmetry is a sort of fermionic generalization (with
anti-commuting parameters) of the geometric invariance at the base
of general relativity (the invariance under any change in
coordinates). The generalization of Einstein's theory of
gravitation that admits such a local supersymmetry is called {\it
supergravity theory}. As we have mentioned, in four dimensions
this theory contains, in addition to the (commuting) bosonic field
$g_{\mu\nu} (x)$, an (anti-commuting) fermionic field $\psi_{\mu}
(x)$ that is both a space-time vector (with index $\mu$) and a
spinor. (It is a massless field of spin $3/2$, intermediate
between a massless spin $1$ field like $A_{\mu}$ and a massless
spin $2$ field like $h_{\mu\nu} = g_{\mu\nu} - \eta_{\mu\nu}$.)
Supergravity was extended to richer fermionic structures (with
many gravitinos), and was formulated in space-times having more
than four dimensions.  It is nevertheless remarkable that there is
a maximal dimension, equal to $D = 11$, admitting a theory of
supergravity (the maximal supergravity constructed by E.~Cremmer,
B.~Julia, and J.~Scherk). The initial hope underlying the
construction of these supergravity theories was that they would
perhaps allow one to give meaning to the perturbative calculation
(\ref{rg23}) of quantum amplitudes. Indeed, one finds for example
that at one loop, $\ell = 1$, the contributions coming from
intermediate fermionic states have a sign opposite to the bosonic
contributions and (because of the supersymmetry, bosons
$\leftrightarrow$ fermions) exactly cancel them. Unfortunately,
although such cancellations exist for the lowest orders of
approximation, it appeared that this was probably not
going to be the case at all orders\footnote{Recent work by
Z. Bern et al. and M. Green et al., has, however, suggested
that such cancellations take place at all orders for the case 
of maximal supergravity, dimensionally reduced to $D=4$ dimensions.}.
 The fact that the
gravitational interaction constant $G$ has ``a bad dimension''
remains true and creates non-renormalizable divergences starting
at a certain number of loops $\ell$.

Meanwhile, a third way of defining a consistent quantum theory of
gravity was developed, under the name of {\it string theory}.
Initially formulated as models for the strong interactions (in
particular by G.~Veneziano, M.~Virasoro, P.~Ramond, A.~Neveu, and
J.H.~Schwarz), the string theories were founded upon the
quantization of the relativistic dynamics of an extended object of
one spatial dimension: a ``string.'' This string could be closed
in on itself, like a small rubber band (a closed string), or it
could have two ends (an open string). Note that the point of
departure of string theory only includes the Poincar\'e-Minkowski
space-time, in other words the metric $\eta_{\mu\nu}$ of
Equation~(\ref{rg2}), and quantum theory (with the constant $\hbar
= h/2\pi$). In particular, the only symmetry manifest in the
classical dynamics of a string is the Poincar\'e group
(\ref{rg3}). It is, however, remarkable that (as shown by
T.~Yoneya, and J.~Scherk and J.H.~Schwarz, in 1974) one of the
quantum excitations of a closed string reproduces, in a certain
limit, all of the non-linear structure of general relativity (see
below). Among the other remarkable properties of string theory
\cite{strings}, let us point out that it is the first physical
theory to determine the space-time dimension $D$. In fact, this
theory is only consistent if $D=10$, for the versions allowing
fermionic excitations (the purely bosonic string theory selects
$D=26$).  The fact that $10 > 4$ does not mean that this theory
has no relevance to the real world. Indeed, it has been known
since the 1930s (from work of T.~Kaluza and O.~Klein) that a
space-time of dimension $D > 4$ is compatible with experiment if
the supplementary (spatial) dimensions close in on themselves
(meaning they are {\it compactified}) on very small distance
scales. The low-energy physics of such a theory seems to take
place in a four-dimensional space-time, but it contains new (a
priori massless) fields connected to the geometry of the
additional compactified dimensions. Moreover, recent work (due in
particular to I.~Antoniadis, N.~Arkani-Hamed, S.~Dimopoulos, and
G.~Dvali) has suggested the possibility that the additional
dimensions are compactified on scales that are small with respect
to everyday life, but very large with respect to the Planck
length. This possibility opens up an entire phenomenological field
dealing with the eventual observation of signals coming from
string theory (see the contribution of I.~Antoniadis to this
Poincar\'e seminar).

However, string theory's most remarkable property is that it seems
to avoid, in a radical way, the problems of divergent
(non-renormalizable) integrals that have weighed down every direct
attempt at perturbatively quantizing gravity. In order to explain
how string theory arrives at such a result, we must discuss some
elements of its formalism.

Recall that the classical dynamics of any system is obtained by
minimizing a functional of the time evolution of the system's
configuration, called the {\it action} (the principle of least
action). For example, the action for a particle of mass $m$,
moving in a Riemannian space-time (\ref{rg6}), is proportional to
the length of the line that it traces in space-time: $S = -m \int
ds$. This action is minimized when the particle follows a
geodesic, in other words when its equation of motion is given by
(\ref{rg7}). According to Y.~Nambu and T.~Goto, the action for a
string is $S = -T \iint dA$, where the parameter $T$ (analogous to
$m$ for the particle) is called the string {\it tension}, and
where $\iint dA$ is the area of the {\it two-dimensional} surface
traced out by the evolution of the string in the ($D$-dimensional)
space-time in which it lives. In quantum theory, the action
functional serves (as shown by R.~Feynman) to define the
transition amplitude (\ref{rg23}). Basically, when one considers
two intermediate configurations $m$ and $n$ (in the sense of the
right-hand side of (\ref{rg23})) that are close to each other, the
amplitude $A(n,m)$ is proportional to $\exp ( i \, S (n,m) /
\hbar)$, where $S(n,m)$ is the minimal classical action such that
the system considered evolves from the configuration labelled by
$n$ to that labelled by $m$. Generalizing the decomposition in
(\ref{rg23}) by introducing an infinite number of intermediate
configurations that lie close to each other, one ends up (in a
generalization of Huygens' principle) expressing the amplitude
$A(f,i)$ as a multiple sum over all of the ``paths'' (in the
configuration space of the system studied) connecting the initial
state $i$ to the final state $f$. Each path contributes a term
$e^{i\phi}$ where the phase $\phi = S/\hbar$ is proportional to
the action $S$ corresponding to this ``path,'' or in other words
to this possible evolution of the system. In string theory, $\phi
= - (T/\hbar) \iint dA$. Since the phase is a non-dimensional
quantity, and $\iint dA$ has the dimension of an area, we see that
the quantum theory of strings brings in the quantity $\hbar / T$,
having the dimensions of a length squared, at a fundamental level.
More precisely, the fundamental length of string theory, $\ell_s$,
is defined by
\begin{equation}
\label{rg25}
\ell_s^2 \equiv \alpha' \equiv \frac{\hbar}{2 \, \pi \, T} \, .
\end{equation}

This fundamental length plays a central role in string theory.
Roughly speaking, it defines the characteristic ``size'' of the quantum states of
a string. If $\ell_s$ is much smaller than the observational
resolution with which one studies the string, the string will look
like a point-like particle, and its interactions will be described
by a quantum theory of relativistic particles, which is equivalent
to a theory of relativistic fields. It is precisely in this sense
that general relativity emerges as a limit of string theory. Since
this is an important conceptual point for our story, let us give
some details about the emergence of general relativity from string
theory.

The action functional that is used in practice to quantize a
string is not really $-T \iint dA$, but rather (as emphasized by
A.~Polyakov)
\begin{equation}
\label{rg26}
\frac{S}{\hbar} = - \frac{1}{4 \, \pi \, \ell_s^2} \iint d^2 \sigma \,
\sqrt{-\gamma} \, \gamma^{ab} \, \partial_a \, X^{\mu} \, \partial_b \, X^{\nu}
\, \eta_{\mu\nu} + \cdots \, ,
\end{equation}
where $\sigma^a$, $a=0,1$ are two coordinates that allow an event
to be located on the space-time surface (or `world-sheet') traced out by the string
within the ambient space-time; $\gamma_{ab}$ is an auxiliary
metric ($d \, \Sigma^2 = \gamma_{ab} (\sigma) \, d\sigma^a \,
d\sigma^b$) defined on this surface (with $\gamma^{ab}$ being its
inverse, and $\gamma$ its determinant); and $X^{\mu} (\sigma^a)$
defines the embedding of the string in the ambient (flat)
space-time. The dots indicate additional terms, and in particular
terms of fermionic type that were introduced by P.~Ramond, by
A.~Neveu and J.H.~Schwarz, and by others. If one separates the two
coordinates $\sigma^a = (\sigma^0 , \sigma^1)$ into a temporal
coordinate, $\tau \equiv \sigma^0$, and a spatial coordinate,
$\sigma \equiv \sigma^1$, the configuration ``at time $\tau$'' of
the string is described by the functions $X^{\mu} (\tau ,
\sigma)$, where one can interpret $\sigma$ as a curvilinear
abscissa describing the spatial extent of the string. If we
consider a closed string, one that is topologically equivalent to
a circle, the function $X^{\mu} (\tau , \sigma)$ must be periodic
in $\sigma$. One can show that (modulo the imposition of certain
constraints) one can choose the coordinates $\tau$ and $\sigma$ on
the string such that $d \, \Sigma^2 = -d\tau^2 + d \sigma^2$.
Then, the dynamical equations for the string (obtained by
minimizing the action (\ref{rg26})) reduce to the standard
equation for waves on a string: $-\partial^2 X^{\mu} /
\partial \tau^2 +
\partial^2 X^{\mu} / \partial \sigma^2 = 0$. The general solution
to this equation describes a superposition of waves travelling
along the string in both possible directions: $X^{\mu} = X_L^{\mu}
(\tau + \sigma) + X_R^{\mu} (\tau - \sigma)$. If we consider a
closed string (one that is topologically equivalent to a circle),
these two types of wave are independent of each other. For an open
string (with certain reflection conditions at the endpoints of the
string) these two types of waves are connected to each other.
Moreover, since the string has a finite length in both cases, one
can decompose the left- or right-moving waves $X_L^{\mu} (\tau +
\sigma)$ or $X_R^{\mu} (\tau - \sigma)$ as a Fourier series. For
example, for a closed string one may write
\begin{equation}
\label{rg27}
X^{\mu} (\tau , \sigma) = X_0^{\mu} (\tau) + \frac{i}{\sqrt 2} \, \ell_s
\sum_{n=1}^{\infty} \left( \frac{a_n^{\mu}}{\sqrt n} \, e^{-2in(\tau - \sigma)}
+ \frac{\tilde a_n^{\mu}}{\sqrt n} \, e^{-2 in (\tau + \sigma)} \right) + {\rm
h.c.}
\end{equation}
Here $X_0^{\mu} (\tau) = x^{\mu} + 2 \, \ell_s^2 \, p^{\mu} \tau$
describes the motion of the string's center of mass, and the
remainder describes the decomposition of the motion around the
center of mass into a discrete set of oscillatory modes. Like any
vibrating string, a relativistic string can vibrate in its
fundamental mode ($n=1$) or in a ``harmonic'' of the fundamental
mode (for an integer $n > 1$). In the classical case the complex
coefficients $a_n^{\mu}$, $\tilde a_n^{\mu}$ represent the
(complex) amplitudes of vibration for the modes of oscillation at
frequency $n$ times the fundamental frequency. (with $a_n^{\mu}$
corresponding to a wave travelling to the right, while $\tilde
a_n^{\mu}$ corresponds to a wave travelling to the left.) When one
quantizes the string dynamics the position of the string $X^{\mu}
(\tau , \sigma)$ becomes an operator (acting in the space of
quantum states of the system), and because of this the quantities
$x^{\mu} , p^{\mu} , a_n^{\mu}$ and $\tilde a_n^{\mu}$ in
(\ref{rg27}) become operators. The notation h.c. signifies that
one must add the hermitian conjugates of the oscillation terms,
which will contain the operators $(a_n^{\mu})^{\dagger}$ and
$(\tilde a_n^{\mu})^{\dagger}$. (The notation $\dagger$ indicates
hermitian conjugation, in other words the operator analog of
complex conjugation.)
 One then finds that the operators $x^{\mu}$ and $p^{\mu}$ describing
 the motion of the center of mass satisfy the usual commutation
 relations of a relativistic particle, $[x^{\mu} , p^{\mu}] = i \,
\hbar \, \eta^{\mu\nu}$, and that the operators $a_n^{\mu}$ and
$\tilde a_n^{\mu}$ become annihilation operators, like those that
appear in the quantum theory of any vibrating system: $[a_n^{\mu}
, (a_m^{\nu})^{\dagger}] = \hbar \, \eta^{\mu\nu} \, \delta_{nm}$,
$[\tilde a_n^{\mu} , (\tilde a_m^{\nu})^{\dagger}] = \hbar \,
\eta^{\mu\nu} \, \delta_{mn}$. In the case of an open string, one
only has {\it one} set of oscillators, let us say $a_n^{\mu}$. The
discussion up until now has neglected to mention that the
oscillation amplitudes $a_n^{\mu} , \tilde a_n^{\mu}$ must satisfy
an infinite number of constraints (connected with the equation
obtained by minimizing (\ref{rg26}) with respect to the auxiliary
metric $\gamma_{ab}$). One can satisfy these by expressing {\it
two} of the space-time components of the oscillators $a_n^{\mu} ,
\tilde a_n^{\mu}$ (for each $n$) as a function of the other.
Because of this, the physical states of the string are described
by oscillators $a_n^i , \tilde a_n^i$ where the index $i$ only
takes $D-2$ values in a space-time of dimension $D$. Forgetting
this subtlety for the moment (which is nevertheless crucial
physically), let us conclude this discussion by summarizing the
{\it spectrum} of a quantum string, or in other words the ensemble
of quantum states of motion for a string.

For an open string, the ensemble of quantum states describes the
states of motion (the momenta $p^{\mu}$) of an infinite collection
of relativistic particles, having squared masses $M^2 = -
\eta_{\mu\nu} \, p^{\mu} \, p^{\nu}$ equal to $(N-1) \, \, m_s^2$,
where $N$ is a non-negative integer and $m_s \equiv \hbar /
\ell_s$ is the fundamental mass of string theory associated to the
fundamental length $\ell_s$. For a closed string, one finds
another ``infinite tower'' of more and more massive particles,
this time with $M^2 = 4 (N-1) \, m_s^2$. In both cases the integer
$N$ is given, as a function of the string's oscillation amplitudes
(travelling to the right), by
\begin{equation}
\label{rg28}
N = \sum_{n=1}^{\infty} n \, \eta_{\mu\nu} (a_n^{\mu})^{\dagger} \, a_n^{\nu} \, .
\end{equation}
In the case of a closed string one must also satisfy the
constraint $N = \tilde N$ where $\tilde N$ is the operator
obtained by replacing $a_n^{\mu}$ by $\tilde a_n^{\mu}$ in
(\ref{rg28}).

The preceding result essentially states that the (quantized)
internal energy of an oscillating string defines the squared mass
of the associated particle. The presence of the additional term
$-1$ in the formulae given above for $M^2$ means that the quantum
state of minimum internal energy for a string, that is, the
``vacuum'' state $\vert 0 \rangle$ where all oscillators are in
their ground state, $a_n^{\mu} \mid 0 \rangle = 0$, corresponds to
a negative squared mass ($M^2 = -m_s^2$ for the open string and
$M^2 = - 4 \, m_s^2$ for the closed string). This unusual quantum
state (a {\it tachyon}) corresponds to an instability of the
theory of bosonic strings. It is absent from the more
sophisticated versions of string theory (``superstrings'') due to
F.~Gliozzi, J.~Scherk, and D.~Olive, to M.~Green and J.H.~Schwarz,
and to D.~Gross and collaborators. Let us concentrate on the other
states (which are the only ones that have corresponding states in
superstring theory). One then finds that the first possible
physical quantum states (such that $N=1$)
 describe some massless particles. In relativistic quantum theory
it is known that any particle is the quantized excitation of a 
corresponding field. Therefore the
massless particles that appear in string theory must correspond to
long-range fields. To know which fields appear in this way one
must more closely examine which possible combinations of
oscillator excitations $a_1^{\mu} , a_2^{\mu} , a_3^{\mu} ,
\ldots$, appearing in Formula (\ref{rg28}), can lead to $N=1$.
Because of the factor $n$ in (\ref{rg28}) multiplying the harmonic
contribution of order $n$ to the mass squared, only the
oscillators of the fundamental mode $n=1$ can give $N=1$. One then
deduces that the internal quantum states of massless particles
appearing in the theory of {\it open strings} are described by a
string oscillation state of the form
\begin{equation}
\label{rg29}
\zeta_{\mu} (a_1^{\mu})^{\dagger} \mid 0 \rangle \, .
\end{equation}
On the other hand, because of the constraint $N = \tilde N = 1$,
the internal quantum states of the massless particles appearing in
the theory of {\it closed strings} are described by a state of
excitation containing both a left-moving oscillation and a
right-moving oscillation:
\begin{equation}
\label{rg30}
\zeta_{\mu\nu} (a_1^{\mu})^{\dagger} \, (\tilde a_1^{\nu})^{\dagger} \mid 0 \rangle \, .
\end{equation}
In Equations (\ref{rg29}) and (\ref{rg30}) the state $\vert 0
\rangle$ denotes the ground state of all oscillators ($a_n^{\mu}
\mid 0 \rangle = \tilde a_n^{\mu} \mid 0 \rangle = 0$).

The state (\ref{rg29}) therefore describes a massless particle
(with momentum satisfying $\eta_{\mu\nu} \, p^{\mu} \, p^{\nu} =
0$), possessing an ``internal structure'' described by a vector
polarization $\zeta_{\mu}$. Here we recognize exactly the
definition of a photon, the quantum state associated with a wave
$A_{\mu} (x) = \zeta_{\mu} \exp (i \, k_{\lambda} \,
x^{\lambda})$, where $p^{\mu} = \hbar \, k^{\mu}$. The theory of
open strings therefore contains Maxwell's theory. (One can also
show that, because of the constraints briefly mentioned above, the
polarization $\zeta_{\mu}$ must be transverse, $k^{\mu} \,
\zeta_{\mu} = 0$, and that it is only defined up to a gauge
transformation: $\zeta'_{\mu} = \zeta_{\mu} + a \, k_{\mu}$.) As
for the state (\ref{rg30}), this describes a massless particle
($\eta_{\mu\nu} \, p^{\mu} \, p^{\nu} = 0$), possessing an
``internal structure'' described by a tensor polarization
$\zeta_{\mu\nu}$. The plane wave associated with such a particle
is therefore of the form $\bar h_{\mu\nu} (x) = \zeta_{\mu\nu}
\exp (i \, k_{\lambda} \, x^{\lambda})$, where $p^{\mu} = \hbar \,
k^{\mu}$. As in the case of the open string, one can show that
$\zeta_{\mu\nu}$ must be transverse, $\zeta_{\mu\nu} \, k^{\nu} =
0$ and that it is only defined up to a gauge transformation,
$\zeta'_{\mu\nu} = \zeta_{\mu\nu} + k_{\mu} \, a_{\nu} + k_{\nu}
\, b_{\mu}$. We here see the same type of structure appear that we
had in general relativity for plane waves. However, here we have a
structure that is richer than that of general relativity. Indeed,
since the state (\ref{rg30}) is obtained by combining two
independent states of oscillation, $(a_1^{\mu})^{\dagger}$ and
$(\tilde a_1^{\mu})^{\dagger}$, the polarization tensor
$\zeta_{\mu\nu}$ is not constrained to be symmetric. Moreover it
is not constrained to have vanishing trace. Therefore, if we
decompose $\zeta_{\mu\nu}$ into its possible irreducible parts (a
symmetric traceless part, a symmetric part with trace, and an
antisymmetric part) we find that the field $\bar h_{\mu\nu} (x)$
associated with the massless states of a closed string decomposes
into: (i) a field $h_{\mu\nu} (x)$ (the {\it graviton})
representing a weak gravitational wave in general relativity, (ii)
a scalar field $\Phi (x)$ (called the {\it dilaton}), and (iii) an
antisymmetric tensor field $B_{\mu\nu} (x) = - B_{\nu\mu} (x)$
subject to the gauge invariance $B'_{\mu\nu} (x) = B_{\mu\nu} (x)
+ \partial_{\mu} \, a_{\nu} (x) -
\partial_{\nu} \, a_{\mu} (x)$. Moreover, when one studies the
non-linear interactions between these various fields, as described
by the transition amplitudes $A(f,i)$ in string theory, one can
show that the field $h_{\mu\nu} (x)$ truly represents a
deformation of the flat geometry of the background space-time in
which the theory was initially formulated. Let us emphasize this
remarkable result. We started from a theory that studied the
quantum dynamics of a string in a rigid background space-time.
This theory predicts that {\it certain quantum excitations of a
string} (that propagate at the speed of light) {\it in fact
represent waves of deformation of the space-time geometry}. In
intuitive terms, the ``elasticity'' of space-time postulated by
the theory of general relativity appears here as being due to
certain internal vibrations of an elastic object extended in one
spatial dimension.

Another suggestive consequence of string theory is the link
suggested by the comparison between (\ref{rg29}) and (\ref{rg30}).
Roughly, Equation~(\ref{rg30}) states that the internal state of a
closed string corresponding to a graviton is constructed by taking
the (tensor) product of the states corresponding to photons in the
theory of open strings. This unexpected link between Einstein's
gravitation $(g_{\mu\nu})$ and Maxwell's theory $(A_{\mu})$
translates, when we look at interactions in string theory, into
remarkable identities (due to H.~Kawai, D.C.~Lewellen, and
S.-H.H.~Tye) between the transition amplitudes of open strings and
those of closed strings. This affinity between electromagnetism,
or rather Yang-Mills theory, and gravitation has recently given
rise to fascinating conjectures (due to A. Polyakov and J.
Maldacena) connecting quantum Yang-Mills theory in flat space-time
to quasi-classical limits of string theory and gravitation in
curved space-time. Einstein would certainly have been interested
to see how classical general relativity is used here to clarify
the limit of a {\it quantum} Yang-Mills theory.

Having explained the starting point of string theory, we can
outline the intuitive reason for which this theory avoids the
problems with divergent integrals that appeared when one tried to
directly quantize gravitation. We have seen that string theory
contains an infinite tower of particles whose masses grow with the
degree of excitation of the string's internal oscillators. The
gravitational field appears in the limit that one considers the
low energy interactions ($E \ll m_s$) between the massless states
of the theory. In this limit the graviton (meaning the particle
associated with the gravitational field) is treated as a
``point-like'' particle. When we consider more complicated
processes (at one loop, $\ell = 1$, see above), virtual elementary
gravitons could appear with arbitrarily high energy. It is these
virtual high-energy gravitons that are responsible for the
divergences. However, in string theory, when we consider any
intermediate process whatsoever where high energies appear, it
must be remembered that this high intermediate energy can also be
used to excite the internal state of the virtual gravitons, and
thus reveal that they are ``made'' from an extended string. An
analysis of this fact shows that string theory introduces an
effective truncation of the type $E \lesssim m_s$ on the energies
of exchanged virtual particles. In other words, the fact that
there are no truly ``point-like'' particles in string theory, but
only string excitations having a characteristic length $\sim
\ell_s$, eliminates the problem of infinities connected to
arbitrarily small length and time scales. Because of this, in
string theory one can calculate the transition amplitudes
corresponding to a collision between two gravitons, and one finds
that the result is given by a finite integral \cite{strings}.

Up until now we have only considered the starting point of string
theory. This is a complex theory that is still in a stage of rapid
development. Let us briefly sketch some other aspects of this
theory that are relevant for this expos\'e centered around
relativistic gravitation. Let us first state that the more
sophisticated versions of string theory ({\it superstrings})
require the inclusion of fermionic oscillators $b_n^{\mu}$,
$\tilde b_n^{\mu}$, in addition to the bosonic oscillators
$a_n^{\mu}$, $\tilde a_n^{\mu}$ introduced above. One then finds
that there are no particles of negative mass-squared, and that the
space-time dimension $D$ must be equal to 10. One also finds that
the massless states contain more states than those indicated
above. In fact, one finds that the fields corresponding to these
states describe the various possible theories of supergravity in
$D=10$. Recently (in work by J.~Polchinski) it has also been
understood that string theory contains not only the states of
excitation of strings (in other words of objects extended in one
spatial direction), but also the states of excitation of objects
extended in $p$ spatial directions, where the integer $p$ can take
other values than $1$. For example, $p=2$ corresponds to a {\it
membrane}. It even seems (according to C.~Hull and P.~Townsend)
that one should recognize that there is a sort of ``democracy''
between several different values for $p$. An object extended in
$p$ spatial directions is called a {\it $p$-brane}. In general,
the masses of the quantum states of these $p$-branes are very
large, being parametrically higher than the characteristic mass
$m_s$. However, one may also consider a limit where the mass of
certain $p$-branes tends towards zero. In this limit, the fields
associated with these $p$-branes become long-range fields. A
surprising result (by E.~Witten) is that, in this limit, the
infinite tower of states of certain $p$-branes (in particular for
$p=0$) corresponds exactly to the infinite tower of states that
appear when one considers the maximal supergravity in $D=11$
dimensions, with the eleventh (spatial) dimension compactified on
a circle (that is to say with a periodicity condition on
$x^{11}$). In other words, in a certain limit, a theory of
superstrings in $D=10$ transforms into a theory that lives in
$D=11$ dimensions! Because of this, many experts in string theory
believe that the true definition of string theory (which is still
to be found) must start from a theory (to be defined) in 11
dimensions (known as ``$M$-theory'').

We have seen in Section~\ref{sec8} that one point of contact
between relativistic gravitation and quantum theory is the
phenomenon of thermal emission from black holes discovered by
S.W.~Hawking. String theory has shed new light upon this
phenomenon, as well as on the concept of black hole ``entropy.''
The essential question that the calculation of S.W.~Hawking left
in the shadows is: what is the physical meaning of the quantity
$S$ defined by Equation~(\ref{rg19})? In the thermodynamic theory
of ordinary bodies, the entropy of a system is interpreted, since
Boltzmann's work, as the (natural) logarithm of the number of
microscopic states $N$ having the same macroscopic characteristics
(energy, volume, etc.) as the state of the system under
consideration: $S = \log N$. Bekenstein had attempted to estimate
the number of microscopic internal states of a macroscopically
defined black hole, and had argued for a result such that $\log N$
was on the order of magnitude of $A / \hbar \, G$, but his
arguments remained indirect and did not allow a clear meaning to
be attributed to this counting of microscopic states. Work by
A.~Sen and by A.~Strominger and C.~Vafa, as well as by C.G.~Callan
and J.M.~Maldacena has, for the first time, given examples of
black holes whose microscopic description in string theory is
sufficiently precise to allow for the calculation (in certain
limits) of the number of internal quantum states, $N$. It is
therefore quite satisfying to find a final result for $N$ whose
logarithm is {\it precisely} equal to the expression (\ref{rg19}).
However, there do remain dark areas in the understanding of the
quantum structure of black holes. In particular, the string theory
calculations allowing one to give a precise statistical meaning to
the entropy (\ref{rg19}) deal with very special black holes (known
as {\it extremal} black holes, which have the maximal electric
charge that a black hole with a regular horizon can support).
These black holes have a Hawking temperature equal to zero, and
therefore do not emit thermal radiation. They correspond to {\it
stable} states in the quantum theory. One would nevertheless also
like to understand the detailed internal quantum structure of {\it
unstable} black holes, such as the Schwarzschild black hole
(\ref{rg17}), which has a non-zero temperature, and which
therefore loses its mass little by little in the form of thermal
radiation. What is the final state to which this gradual process
of black hole ``evaporation'' leads? Is it the case that an
initial pure quantum state radiates all of its initial mass to
transform itself entirely into incoherent thermal radiation? Or
does a Schwarzschild black hole transform itself, after having
obtained a minimum size, into something else? The answers to these
questions remain open to a large extent, although it has been
argued that a Schwarzschild black hole transforms itself into a
highly massive quantum string state when its radius becomes on the
order of $\ell_s$ \cite{bh}.

We have seen previously that string theory contains general
relativity in a certain limit. At the same time, string theory is,
strictly speaking, infinitely richer than Einstein's gravitation,
for the graviton is nothing more than a particular quantum
excitation of a string, among an infinite number of others. What
deviations from Einstein's gravity are predicted by string theory?
This question remains open today because of our lack of
comprehension about the connection between string theory and the
reality observed in our everyday environment (4-dimensional
space-time; electromagnetic, weak, and strong interactions; the
spectrum of observed particles; $\ldots$). We shall content
ourselves here with outlining a few possibilities. (See the
contribution by I.~Antoniadis for a discussion of other
possibilities.) First, let us state that if one considers
collisions between gravitons with energy-momentum $k$ smaller
than, but not negligible with respect to, the characteristic
string mass $m_s$, the calculations of transition amplitudes in
string theory show that the usual Einstein equations (in the
absence of matter) $R_{\mu\nu} = 0$ must be modified, by including
corrections of order $(k/m_s)^2$. One finds that these modified
Einstein equations have the form (for bosonic string theory)
\begin{equation}
\label{rg31}
R_{\mu\nu} + \frac{1}{4} \, \ell_s^2 \, R_{\mu\alpha\beta\gamma} \,
R_{\nu}^{\centerdot\alpha\beta\gamma} + \cdots = 0 \, ,
\end{equation}
where
\begin{equation}
\label{rg32}
R_{\centerdot\nu\alpha\beta}^{\mu} \equiv \partial_{\alpha} \,
\Gamma_{\nu\beta}^{\mu} + \Gamma_{\sigma\alpha}^{\mu} \,
\Gamma_{\nu\beta}^{\sigma} - \partial_{\beta} \, \Gamma_{\nu\alpha}^{\mu} -
\Gamma_{\sigma\beta}^{\mu} \, \Gamma_{\nu\alpha}^{\sigma} \, ,
\end{equation}
denotes the ``curvature tensor'' of the metric $g_{\mu\nu}$. (the
quantity $R_{\mu\nu}$ defined in Section~\ref{sec5} that appears
in Einstein's equations in an essential way is a ``trace'' of this
tensor: $R_{\mu\nu} = R_{\centerdot\mu\sigma\nu}^{\sigma}$.) As
indicated by the dots in (\ref{rg31}), the terms written are no
more than the two first terms of an infinite series in growing
powers of $\ell_s^2 \equiv \alpha'$. Equation~(\ref{rg31}) shows
how the fact that the string is not a point, but is rather
extended over a characteristic length $\sim \ell_s$, modifies the
Einsteinian description of gravity. The corrections to Einstein's
equation shown in (\ref{rg31}) are nevertheless completely
negligible in most applications of general relativity. In fact, it
is expected that $\ell_s$ is on the order of the Planck scale
$\ell_p$, Equation~(\ref{rg24}). More precisely, one expects that
$\ell_s$ is on the order of magnitude of $10^{-32}$ cm.
(Nevertheless, this question remains open, and it has been
recently suggested that $\ell_s$ is much larger, and perhaps on
the order of $10^{-17}$ cm.)

If one assumes that $\ell_s$ is on the order of magnitude of
$10^{-32}$ cm (and that the extra dimensions are compactified on
distances scales on the order of $\ell_s$), the only area of
general relativistic applications where the modifications shown in
(\ref{rg31}) should play an important role is in primordial
cosmology. Indeed, close to the initial singularity of the Big
Bang (if it exists), the ``curvature'' $R_{\mu\nu\alpha\beta}$
becomes extremely large. When it reaches values comparable to
$\ell_s^{-2}$ the infinite series of corrections in (\ref{rg31})
begins to play a role comparable to the first term, discovered by
Einstein. Such a situation is also found in the interior of a
black hole, when one gets very close to the singularity (see
Figure~\ref{fig3}). Unfortunately, in such situations, one must
take the infinite series of terms in (\ref{rg31}) into account, or
in other words replace Einstein's description of gravitation in
terms of a {\it field} (which corresponds to a {\it point-like}
(quantum) particle) by its exact stringy description. This is a
difficult problem that no one really knows how to attack today.

However, a priori string theory predicts more drastic low energy
($k \ll m_s$) modifications to general relativity than the
corrections shown in (\ref{rg31}). In fact, we have seen in
Equation~(\ref{rg30}) above that Einsteinian gravity does not
appear alone in string theory. It is always necessarily
accompanied by other long-range fields, in particular a scalar
field $\Phi (x)$, the {\it dilaton}, and an antisymmetric tensor
$B_{\mu\nu} (x)$. What role do these ``partners'' of the graviton
play in observable reality? This question does not yet have a
clear answer. Moreover, if one recalls that (super)string theory
must live in a space-time of dimension $D=10$, and that it
includes the $D=10$ (and eventually the $D=11$) theory of
supergravity, there are many other supplementary fields that add
themselves to the ten components of the usual metric tensor
$g_{\mu\nu}$ (in $D=4$). It is conceivable that all of these
supplementary fields (which are massless to first approximation in
string theory) acquire masses in our local universe that are large
enough that they no longer propagate observable effects over
macroscopic scales. It remains possible, however, that one or
several of these fields remain (essentially) massless, and
therefore can propagate physical effects over distances that are
large enough to be observable. It is therefore of interest to
understand what physical effects are implied, for example, by the
dilaton $\Phi (x)$ or by $B_{\mu\nu} (x)$. Concerning the latter,
it is interesting to note that (as emphasized by A.~Connes,
M.~Douglas, and A.~Schwartz), in a certain limit, the presence of
a background $B_{\mu\nu} (x)$ has the effect of deforming the
space-time geometry in a ``non-commutative'' way. This means that,
in a certain sense, the space-time coordinates $x^{\mu}$ cease to
be simple real (commuting) numbers in order to become
non-commuting quantities: $x^{\mu} x^{\nu} - x^{\nu} x^{\mu} =
\varepsilon^{\mu\nu}$ where $\varepsilon^{\mu\nu} = -
\varepsilon^{\nu\mu}$ is connected to a (uniform) background
$B_{\mu\nu}$. To conclude, let us consider the other obligatory
partner of the graviton $g_{\mu\nu} (x)$, the dilaton $\Phi (x)$.
This field plays a central role in string theory. In fact, the
average value of the dilaton (in the vacuum) determines the string
theory coupling constant, $g_s = e^{\Phi}$. The value of $g_s$ in
turn determines (along with other fields) the physical coupling
constants. For example, the gravitational coupling constant is
given by a formula of the type $\hbar \, G = \ell_s^2 (g_s^2 +
\cdots)$ where the dots denote correction terms (which can become
quite important if $g_s$ is not very small). Similarly, the fine
structure constant, $\alpha = e^2 / \hbar c \simeq 1/137$, which
determines the intensity of electromagnetic interactions is a
function of $g_s^2$. Because of these relations between the
physical coupling constants and $g_s$ (and therefore the value of
the dilaton; $g_s = e^{\Phi}$), we see that if the dilaton is
massless (or in other words is long-range), its value $\Phi (x)$
at a space-time point $x$ will depend on the distribution of
matter in the universe. For example, as is the case with the
gravitational field (for example $g_{00} (x) \simeq -1 + 2 GM /
c^2 r$), we expect that the value of $\Phi (x)$ depends on the
masses present around the point $x$, and should be different at
the Earth's surface than it is at a higher altitude. One may also
expect that $\Phi (x)$ would be sensitive to the expansion of the
universe and would vary over a time scale comparable to the age of
the universe. However, if $\Phi (x)$ varies over space and/or
time, one concludes from the relations shown above between $g_s =
e^{\Phi}$ and the physical coupling constants that the latter must
also vary over space and/or time. Therefore, for example, the
value, here and now, of the fine structure constant $\alpha$ could be slightly
different from the value it had, long ago, in a very distant galaxy.
 Such effects are
accessible to detailed astronomical observations and, in fact,
some recent observations have suggested that the interaction
constants were different in distant galaxies. However, other
experimental data (such as the fossil nuclear reactor at Oklo and
the isotopic composition of ancient terrestrial meteorites) put
very severe limits on any variability of the coupling
``constants.'' Let us finally note that if the fine structure
``constant'' $\alpha$, as well as other coupling ``constants,''
varies with a massless field such as the dilaton $\Phi (x)$, then
this implies a violation of the basic postulate of general
relativity: the principle of equivalence. In particular, one can
show that the universality of free fall is necessarily violated,
meaning that bodies with different nuclear composition would fall
with different accelerations in an external gravitational field.
This gives an important motivation for testing the principle of
equivalence with greater precision. For example, the MICROSCOPE
space mission \cite{micro} (of the CNES) should soon test the
universality of free fall to the level of $10^{-15}$, and the STEP
space project (Satellite Test of the Equivalence Principle)
\cite{step} could reach the level $10^{-18}$.

Another interesting phenomenological possibility is that the
dilaton (and/or other scalar fields of the same type, called {\it
moduli}) acquires a non-zero mass that is however very small with
respect to the string mass scale $m_s$. One could then observe a
modification of Newtonian gravitation over small distances
(smaller than a tenth of a millimeter). For a discussion of this
theoretical possibility and of its recent experimental tests see,
respectively, the contributions by I.~Antoniadis and J.~Mester to
this Poincar\'e seminar.

\section{Conclusion}\label{sec12}

For a long time general relativity was admired as a marvellous
intellectual construction, but it only played a marginal role in
physics.  Typical of the appraisal of this theory is the comment
by Max Born \cite{B56} made upon the fiftieth anniversary of the
{\it annus mirabilis}: ``The foundations of general relativity
seemed to me then, and they still do today, to be the greatest
feat of human thought concerning Nature, the most astounding
association of philosophical penetration, physical intuition, and
mathematical ability. However its connections to experiment were
tenuous. It seduced me like a great work of art that should be
appreciated and admired from a distance.''

Today, one century after the {\it annus mirabilis}, the situation
is quite different. General relativity plays a central role in a
large domain of physics, including everything from primordial
cosmology and the physics of black holes to the observation of
binary pulsars and the definition of international atomic time. It
even has everyday practical applications, via the satellite
positioning systems (such as the GPS and, soon, its European
counterpart Galileo). Many ambitious (and costly) experimental
projects aim to test it (G.P.B., MICROSCOPE, STEP,~$\ldots$), or
use it as a tool for deciphering the distant universe
(LIGO/VIRGO/GEO, LISA,~$\ldots$). The time is therefore long-gone
that its connection with experiment was tenuous. Nevertheless, it
is worth noting that the fascination with the structure and
physical implications of the theory evoked by Born remains intact.
One of the motivations for thinking that the theory of strings
(and other extended objects) holds the key to the problem of the
unification of physics is its deep affinity with general
relativity. Indeed, while the attempts at ``Grand Unification''
made in the 1970s completely ignored the gravitational
interaction, string theory necessarily leads to Einstein's
fundamental concept of a dynamical space-time. At any rate, it
seems that one must more deeply understand the 
``generalized quantum geometry'' created through the interaction
of strings and $p$-branes in order to completely formulate this
theory and to understand its hidden symmetries and physical
implications. Einstein would no doubt appreciate seeing the
key role played by symmetry principles and gravity within modern
physics.

\end{document}